\documentclass[aps,prb,twocolumn]{revtex4-1}
\pdfoutput=1
\usepackage{amsmath}
\usepackage{ amssymb }
\usepackage{graphicx}
\usepackage{color}

\begin{document}

\newcommand{\beq}{\begin{equation}}
\newcommand{\eeq}{\end{equation}}
\newcommand{\beqa}{\begin{eqnarray}}
\newcommand{\eeqa}{\end{eqnarray}}
\newcommand{\note}[1]{{\color{red} [#1]}}
\newcommand{\bra}[1]{\ensuremath{\langle#1|}}
\newcommand{\ket}[1]{\ensuremath{|#1\rangle}}
\newcommand{\bracket}[2]{\ensuremath{\langle#1|#2\rangle}}
\renewcommand{\vec}[1]{\textbf{#1}}
\newcommand{\dagga}{{\phantom{\dagger}}}


\title{Parameter diagnostics of   phases and  phase transition learning by neural networks}

\author{Philippe Suchsland}
\affiliation{Institut f\"ur Theoretische Festk\"orperphysik, JARA-FIT and JARA-HPC, RWTH Aachen University, 52056 Aachen, Germany}

\author{Stefan Wessel}
\affiliation{Institut f\"ur Theoretische Festk\"orperphysik, JARA-FIT and JARA-HPC, RWTH Aachen University, 52056 Aachen, Germany}

\date{\today}

\begin{abstract}
We present  an analysis of  neural network-based machine learning schemes for phases and phase transitions in theoretical  condensed matter research, focusing on neural networks with a single  hidden layer. Such shallow neural networks were previously found to be  efficient in classifying phases and locating phase transitions  of various basic model systems. 
In order to rationalize the emergence of the classification process and for identifying any underlying physical quantities, it is  feasible to  examine the weight matrices and the convolutional filter kernels that result from the learning process of such shallow networks. Furthermore, we demonstrate how the learning-by-confusing scheme  can be used, in combination with a simple threshold-value classification method, to diagnose the learning parameters of neural networks. 
In particular, we study the classification process of both fully-connected and convolutional neural networks
for the two-dimensional Ising model with extended domain wall configurations included in the low-temperature regime. Moreover, we consider   the two-dimensional XY model and contrast the performance of the  learning-by-confusing scheme  and  convolutional neural networks trained on bare spin configurations to the case of  preprocessed samples with respect to  vortex configurations. 
We  discuss these findings in relation to  similar  recent investigations and possible further applications. 
\end{abstract}

\maketitle

\section{Introduction}\label{sec:intro}

The use of machine learning approaches is  in  focus of several recent developments
in theoretical condensed matter research. In particular, 
neural networks have been suggested for identifying phases of matter as well as 
phase transitions~\cite{Carrasquilla17,Nieuwenburg17, Hu17, Wetzel17a,Wetzel17b,Ponte17,Liu17, Zhang17, Wang17,Broecker17,Morningstar17,Beach17}.
One motivation behind such proposals is the ability of appropriately designed and trained neural networks, as universal function approximators~\cite{Cybenko89,Hornik91}, 
 to identify patterns from a large set of data~\cite{Nielsen15}. 
In  applications from condensed matter theory,
such  data  sets may consist of sample configurations of a many-body system generated, e.g.,  
by Monte Carlo simulations.  A key point  of such machine learning approaches would be  to minimize the amount of 
preprocessing of the bare sample configurations before feeding them into the learning process,  and to thus 
leave it to the network  to identify the physically relevant  features.
As such, neural networks would indeed be useful computational tools to detect  unexplored phases or phase transitions in condensed matter. 

Within a setting known as supervised learning, the neural network is trained  to distinguish  different phases of a many-body system, based on a large number of training set configurations in combination with appropriate learning schemes. The network's internal classification  should then allow it to associate a previously unseen configuration to the appropriate phase with a high fidelity. 
Consider for example a system that exhibits two different thermodynamic phases, which are separated by a thermal phase transition at a transition temperature $T_c$. Given the value of $T_c$, 
one can explicitly label each  training batch configuration as  belonging to either the high or the low temperature 
phase. Depending on the neural network design, rather high accuracies can  indeed  be  achieved by such supervised learning approaches in  classifying a new configuration as belonging to the high or the low temperature phase~\cite{Carrasquilla17}.

For cases where the actual value of $T_c$ is not known, various schemes have been proposed that use neural networks  to obtain an estimate for $T_c$. 
In one such approach, the confusing scheme~\cite{Nieuwenburg17}, 
 the neural network's ability to identify patterns 
is combined with the idea of labeling the  training batch configurations based on a guess value   of the true $T_c$. The  final estimate for $T_c$  is then obtained as the guess value, for which the network shows an optimal  test accuracy.  Further variants of this semi-unsupervised learning scheme have   been suggested  recently~\cite{Liu17,Broecker17}.  

In general, various neural network designs can be considered for such classification tasks. Deep learning, wherein the neural network exhibits a hierarchy 
of several internal layers  is particularly  prominent for  a broad range of applications~\cite{Goodfellow17}. On the other hand,
for various applications considered in the condensed physics context of phase transitions and the identification of 
phases of matter~\cite{Carrasquilla17,Nieuwenburg17, Hu17, Wetzel17a,Wetzel17b,Ponte17,Liu17, Zhang17, Wang17,Broecker17,Morningstar17,Beach17},
it appears that also networks with only a few hidden layers perform rather well. In contrast to the complexity of  deep learning networks, 
for such shallow networks the classification mechanism resulting from the training phase may still be 
rationalized to a satisfactory degree upon examining the network's connection weights and filter kernels.
As a  further diagnostic tool to identify physical parameters relevant for the classification process, 
we  perform a direct comparison of the network's classification performance to a simple threshold-value classification based on  specific physical observables (we will introduce this approach in Sec.~\ref{Sec:Confusion}). As we will show below, such  diagnostic approaches can  provide  insights into  how the neural network classification actually comes about in a given specific application. 
In the following, we use the basic examples of the  two-dimensional Ising and XY models to perform such a diagnostic analysis of different neural network-based learning schemes. We consider several issues that 
may appear 
in attempts to employ such machine learning methods to study many-body phases and phase transitions of more complex systems.  

The remainder of this paper is organized as follows: In the first part, we concentrate on the Ising model. In particular, 
in Sec.~\ref{Sec:Ising}, we  review the supervised learning approach based on the most basic,  fully connected neural network with a single hidden layer. Here, we furthermore examine in detail  the classification process  for a small number of neurons on the hidden layer.
Then, in Sec.~\ref{Sec:Domainwalls}, we analyze how the   classification process differs,  once a significant amount of low-temperature configurations are included that contain extended domain walls  (EDW).
In such EDW configurations, the system is divided in two oppositely ordered extended domains, along with  domain walls that typically extend across the full linear  size of the system (the inset of Fig.~\ref{fig_dw_fc_N4Tm} provides an example). 
To  correctly classify such EDW configurations,  convolutional neural networks (CNN) are more effective and  we examine in detail, how the different filters contribute to the overall  classification procedure of a CNN in the presence of EDW configurations. 
In the second part, we then consider the XY model. 
First, in Sec.~\ref{Sec:XY}, we examine the classification process of a CNN for the XY model in the context of supervised learning.
In Sec.~\ref{Sec:Confusion}, we then examine the learning-by-confusion scheme as applied to the XY model  and also introduce a threshold-value classification scheme that  turns out to be useful in order to understand both the 
behavior of the CNN as well as the learning-by-confusion scheme for the XY model. We expect this approach to be of value  for the diagnostics of other models,  neural network designs, and  deep learning schemes as well.
Finally, in Sec.~\ref{Sec:Discussion}  we provide a summary of our findings and a comparison to related recent work.

\section{Supervised learning  the Ising model }
\label{Sec:Ising}
The two-dimensional Ising model has been considered  early on in the application of machine learning methods in condensed matter physics. Here, we   revisit in particular the supervised learning approach for classifying  Ising model configurations into the high and low temperature phases by using a simple fully connected feed forward neural network~\cite{Carrasquilla17}.  For this purpose, we feed to  the input layer the real space spin configurations of an Ising model with $N=L\times L$ spins, $\sigma_j=\pm 1, j=1,...,N$, for a finite square lattice with periodic boundary conditions, described by the Hamiltonian
\begin{equation}
H_I=- J \sum_{\langle j,j' \rangle} \sigma_j \sigma_{j'},
\end{equation}
where we fix units in terms of the nearest-neighbor coupling $J=1$.
For the analysis in this section, we  generated spin configurations at different temperatures, weighted by the Boltzmann distribution, using the Wolff cluster algorithm~\cite{Wolff89}. This  allowed us to efficiently generate a large number of uncorrelated sample configurations over a wide temperature range. 
Furthermore, in the low-temperature regime,  the spin configurations obtained this way do not exhibit  extended domain walls (EDW), i.e.,  the obtained low-temperature configurations are strongly polarized either up or down. This turns out to be important for the  learning process. In the next section, we analyze to what extend the learning process differs,  
once a significant number of EDW configurations are present, e.g., when local spin updates are used instead to generate the spin configurations. 

In the following,  we denote the signal of the $j$-th input neuron by $x_j$, such that for a given spin configuration, $x_j=\sigma_j$. 
The fully connected neural network  consists of a single hidden layer with a number $N_f$ of neurons. The activity $z_i$ of its  $i$-th neuron is obtained upon applying an appropriate  activation function~\cite{Nielsen15, Glorot11} $h$ (such as the rectified linear unit ReLU$(\cdot)=\max(0,\cdot)$ or the sigmoidal function $\sigma(\cdot) =[1+1/\exp(\cdot)]^{-1})$ to its
activation $a_i$, i.e. the 
 linearly weighted summation of the input signal, so that
$
 z_i=h(a_i)$, with $a_i=\sum_{j} w_{i,j} x_j +b_i,
 $
 with an $N_f\times N$ weight matrix $w_{i,j}$ and the local bias $b_i$. 
 The activity on the second, output layer with two neurons  is $y_l=\mathrm{softmax}(a'_1, a'_2)$, where $a'_l=\sum_{i} w'_{l,i} z_i +b'_l$, for $l=1,2$,
 in terms of the softmax activation function~\cite{Nielsen15}, again with the corresponding weight matrix and local biases. The ratio $R$ of the two output activities is then obtained as 
 \begin{equation}\label{eq:outratio}
 R=\frac{y_1}{y_2}=e^{b'_1-b'_2} \prod_i e^{z_i (w'_{1,i}-w'_{2,i})},
 \end{equation} 
In the following, the two output neurons correspond to the high ($l=1$) and low ($l=2$) temperature phase, respectively.
Hence, $R$ gives the ratio of the assigned probabilities for classifying a given input configuration to the high- or low-temperature phase.  
The classification task thus  essentially requires learning a threshold-value for the activities $z_i$, as the above ratio is strictly monotone in every argument $z_i$, which are nonnegative numbers.

In previous studies, it was demonstrated that already a narrow hidden layer of  only three ($N_f=3$) neurons  exhibits a high overall classification accuracy~\cite{Carrasquilla17}. The neural network was furthermore found to rely on the  magnetization $m=\frac{1}{N}\sum_j \sigma_j$ of the input configurations to perform the classification.  Namely after training, each of the activations of the hidden units, $a_i$,   showed an essentially linear dependence  on the magnetization $m$ of the input configuration, such that the neural network could be said to have learned the magnetization. 
A network with $N_f=100$ neurons on the hidden layer exhibited a similar behavior. These neurons were of four characteristic types, being active either if the input configuration is dominantly polarized up (or down), or being active if the input states is either polarized up (or down) or unpolarized~\cite{Carrasquilla17}. 
One may  indeed expect such an essentially equal distribution of  a large number of hidden  neurons among  the  low number of different characteristic types  to result from the learning process with random initialization of all  weights and biases. 
\begin{figure}[t]
\includegraphics[width=\columnwidth]{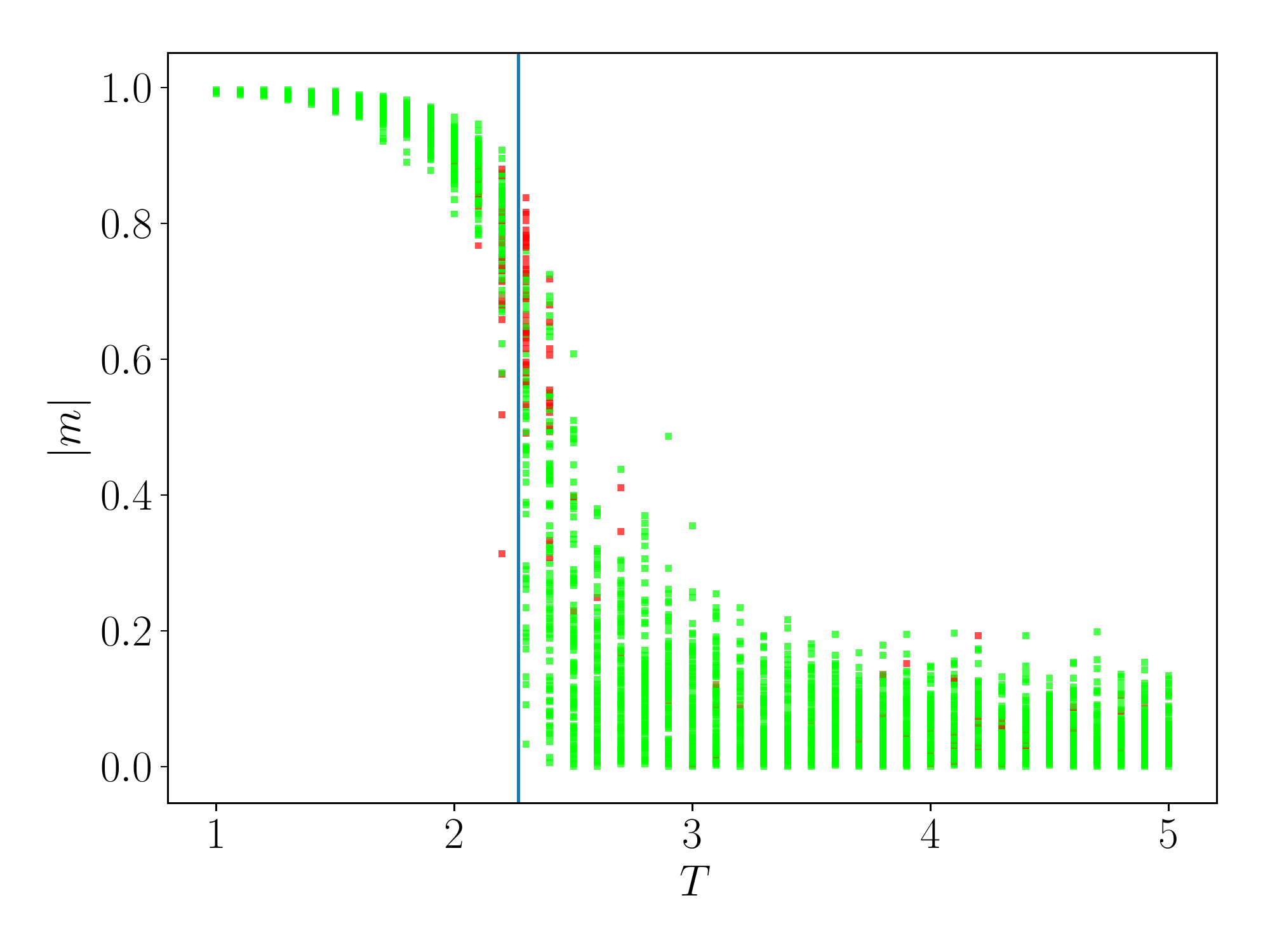}
\caption{(Color online).
Classification correctness  of the $N_f=4$ network  in the temperature $T$ vs.  magnetization $|m|$ plane of the input configurations. Green (red) dots correspond to a correct (wrong) classifcation. The vertical line denotes the exact  transition temperature $T_c$. 
}
\label{fig_fc_N4Tm}
\end{figure}

For a smaller number of hidden neurons, a  less symmetric distributions of the hidden neurons
among the four above types can  arise. Here, we examine  the case of a neural network with $N_f=4$ hidden neurons and  sigmoidal activation, which shows a classification accuracy of  98\% after training on a set of $50000$ configurations for an Ising model with $L=32$ at  temperatures between $T=1$ and $T=5$ with a spacing of $\Delta T=0.1$.
The resulting classification correctness  of the neural network in the plane of the magnetization $|m|$ vs. the temperature $T$ after training is shown in Fig.~\ref{fig_fc_N4Tm}.
For the training phase, we used the Adam method with a cross entropy cost function, and  for larger networks also applied L2-regularization to avoid overfitting~\cite{Kingma14,Nielsen15,Goodfellow17}. All calculations were furthermore performed based on tensorflow~\cite{Tensorflow}.

\begin{figure}[t]
\includegraphics[width=0.9\columnwidth]{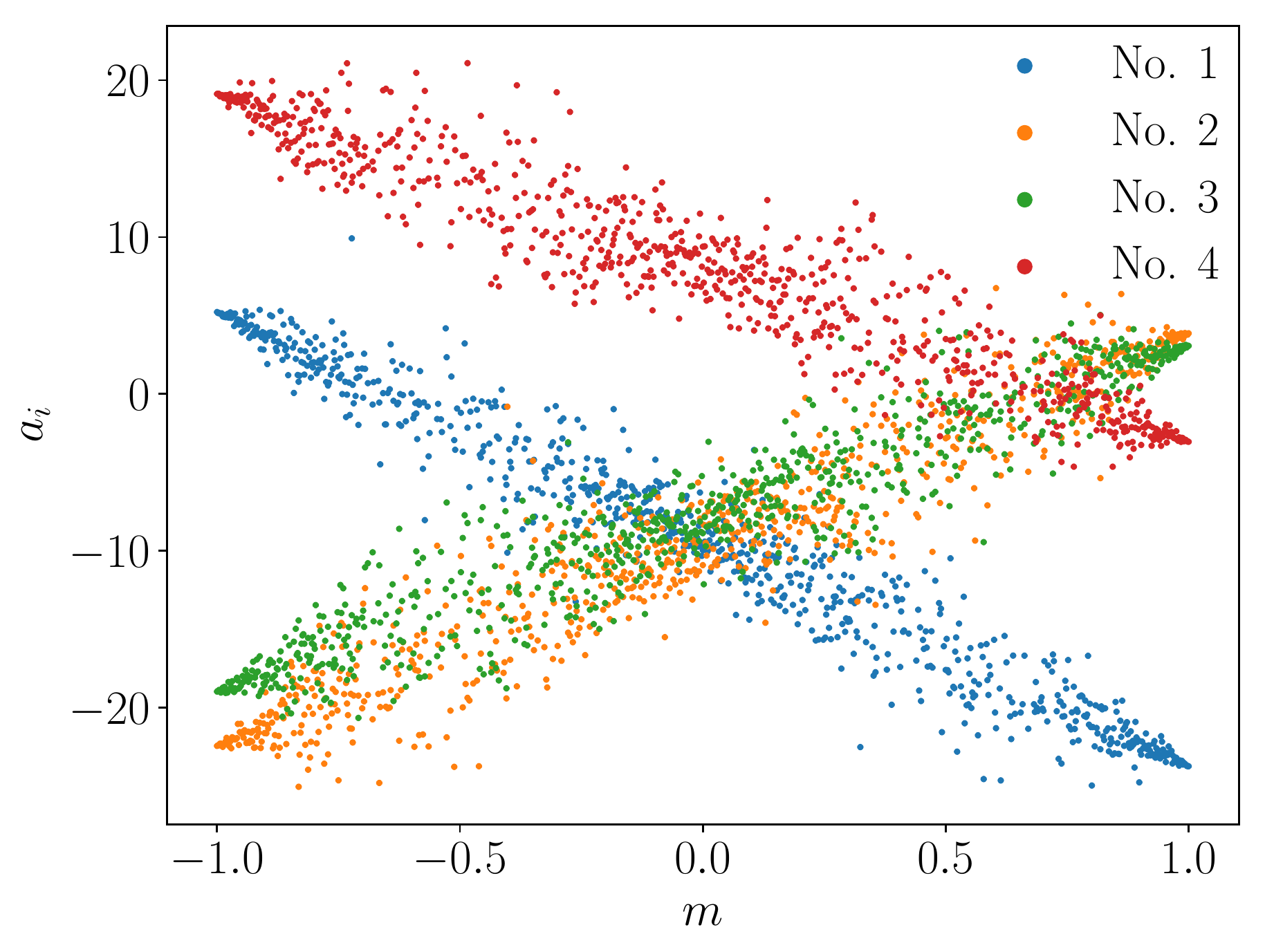}
\caption{(Color online). 
Hidden layer neuron activations $a_i$ for each of the four neurons $i=1,...,4$ of the $N_f=4$ network as a function of the magnetization $m$ of the input configuration.
}
\label{fig_fc_N4a}
\end{figure}

For the $N_f=4$ network, Fig.~\ref{fig_fc_N4a} shows the activations $a_i$ for the hidden layer neurons as a function of the magnetization $m$ of the input configuration:
For this specific network, one neuron (labeled No.~1) apparently  activates for configurations that are dominantly polarized down,  two neurons (No.~2 and 3)  activate for configurations that are  dominantly polarized up, and one neuron (No.~4)  activates for configurations that are either unpolarized or dominantly polarized down.  
The sigmoidal activation is beneficial for obtaining  such linear relations between the magnetization and the activations: 
the  neurons deactivate in one of the two cases of low-temperature polarization, while in the other case
their activity is limited, since the sigmoidal function converges, in contrast to, e.g., the ReLU activation function. 

\begin{figure}[t]
\includegraphics[width=0.9\columnwidth]{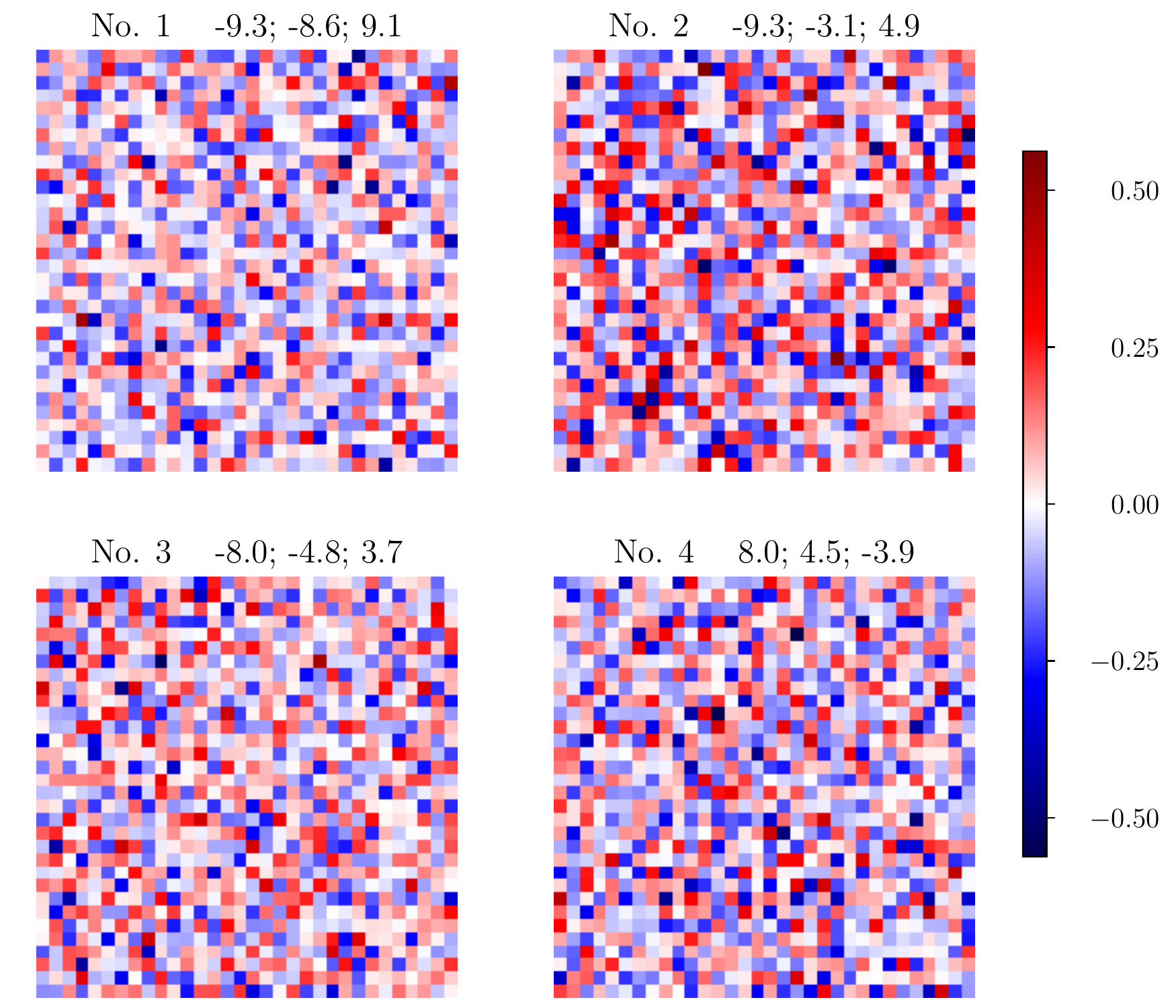}
\caption{(Color online). 
Weight matrices $w_{i,j}$ of the $N_f=4$ network. For each neuron (label by its index $i=1,..,4$), the weight matrix is shown as a two-dimensional array of $32\times 32$ values, corresponding to the layout of the  Ising model configurations. 
The numbers to the right of each neuron $i$ denote the local bias $b_i$, and the weights $w'_{1,i}$, $w'_{2,i}$ that connect this neuron to the high- and low-temperature neuron of the  output layer (in this order). } 
\label{fig_fc_N4w}
\end{figure}

The  final weight matrices $w_{i,j}$ of  the above network  are shown in Fig.~\ref{fig_fc_N4w}, along with the values of the  local biases $b_i$, and the weights $w'_{1,i}$, $w'_{2,i}$, which  connect neuron $i$ to the output layer. 
The featureless noise  in the weights  $w_{i,j}$ reflects  the translational invariance of the Ising model. 
From the signs of the weights  $w'_{l,j}$, we can furthermore identify neurons No. 1, 2, and  3 as  low-temperature activating, while neuron No.~4 is a high-temperature activating neuron: 
this neuron has a positive bias, so that for high-temperature (i.e., disordered) input configurations, for which $\sum_j w_{ij} x_j\approx 0$, this positive bias leads to its activation, which then 
contributes positively to the activation of the high-temperature output neuron. For low temperature input configurations, the color-coded weights in Fig.~\ref{fig_fc_N4w} show a slight  preference for neuron No. 1 towards negative net polarization (blue) and for neurons No.~2 and 3 towards positive net polarization (red). This leads to the $m$-dependence observed in Fig.~\ref{fig_fc_N4a}. Given similar local biases, the approximately twice as large value of the weights  to the output layer for neuron 1 compared to neuron 4  ensure that the low-temperature output neuron's activation is enhanced for dominantly negatively polarized input configuration. 

On the other hand, for low-temperature input configurations with a positive polarization, the fact that two low-temperature active neurons (No.~2 and 3) are present ensures an enhanced activation of 
the low-temperature output neuron, even though the weights to the output layer for neurons No.~2, 3 and 4 are of similar magnitude. We observed such a  balance between the output layer weights and the number of neurons of a specific type to emerge from the learning process also in other trained fully connected networks with a low number of hidden layer neurons. 

\section{Including extended domain wall configurations}
\label{Sec:Domainwalls}
\begin{figure}[t]
\includegraphics[width=\columnwidth]{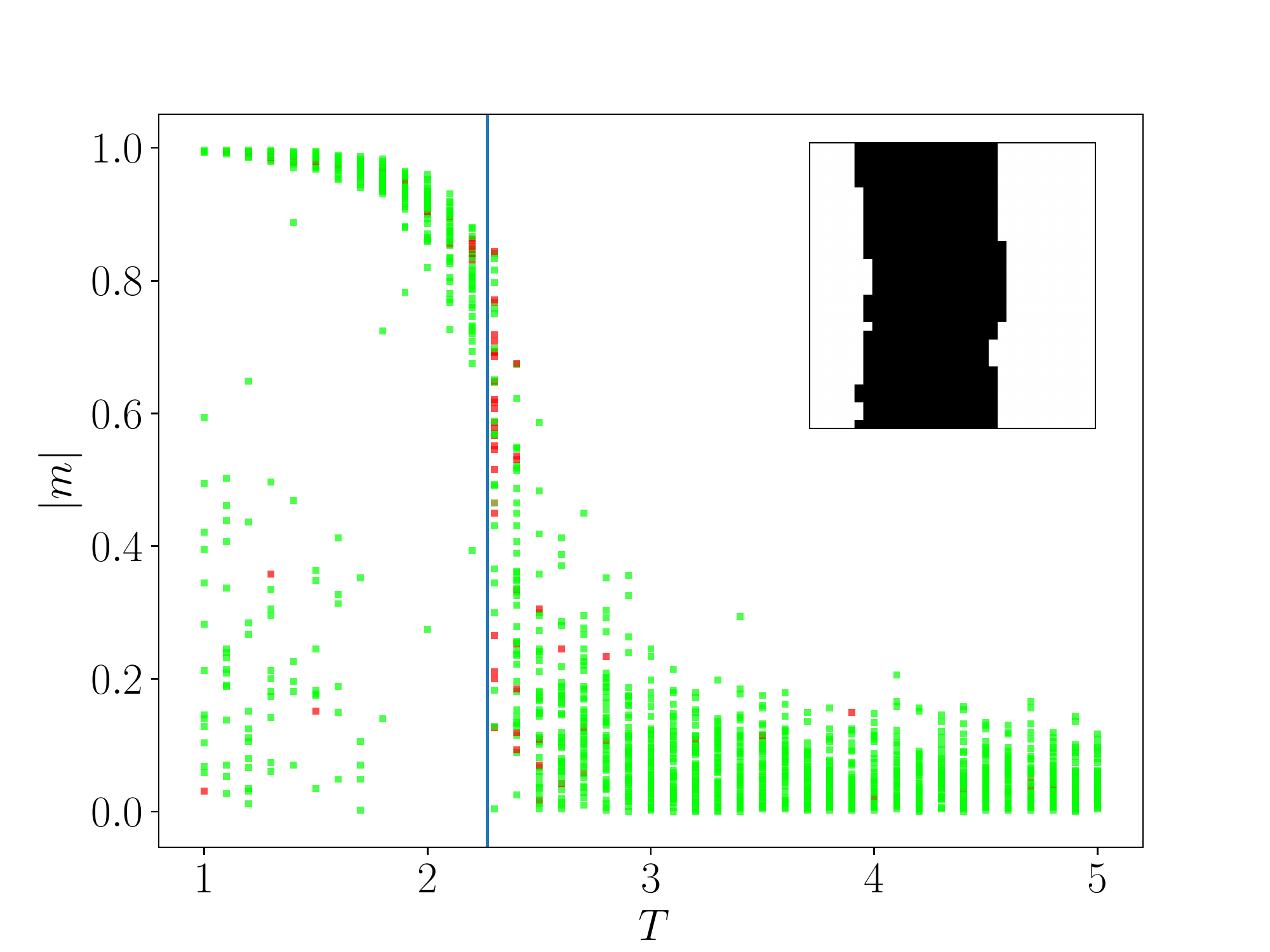}
\caption{(Color online).
Classification correctness  of the $N_f=4$ network for the Ising model with EDW  configurations included, in the temperature $T$ vs.  magnetization $|m|$ plane of the input configurations. Green (red) dots indicate correct (wrong) classifcations. The vertical line denotes the exact  transition temperature $T_c$.  The inset shows a typical EDW configuration with two vertical EDWs  separating  two oppositely polarized spatial regions of similar size on the  $L=32$  lattice. 
}
\label{fig_dw_fc_N4Tm}
\end{figure}

In the supervised learning approach for the Ising model that we reviewed in the previous section, the magnetization $m$  played a crucial  role for the classification task. Being the order parameter,  $m$ is  indeed  a natural quantity  that allows the network to distinguish between configurations from the high- and low-temperature regime. With respect to applications of such machine learning approaches to more complex physical situations, one may thus ask, how the neural network  performs, if the order parameter is not   directly accessible from the input configurations. How will the learning and classification task proceed under such circumstances? In the case of the Ising model, a simple means of eliminating the direct access to the order parameter is to include EDW configurations in the low-temperature regime, such as the configuration shown in the inset of Fig.~\ref{fig_dw_fc_N4Tm} for an  $L=32$  lattice.  Such  EDW configurations often appear in Monte Carlo simulations based on local updates, and can persist  over extended simulation time scales. In more complex systems, such EDW configurations may  be  unavoidable,  for example, if only local update schemes are  available, or if thermalization is not fully controlled. How does the neural network perform the classification task under such conditions? 

Here, we examine this question for the case of the Ising model, using  the Metropolis local spin flip algorithm to generate both  learning and validation configurations.  
We consider again a system with $L=32$ within the same temperature range as in Sec.~\ref{Sec:Ising}. 
It goes without saying that the network  from the  previous section  fails completely to correctly classify any of the low-temperature EDW configurations. 
This is to be expected,  as during the training
period this network was not exposed to any such configurations. 
We thus  consider next  a neural network that was trained on a  data set that includes EDW configurations. 
Figure~\ref{fig_dw_fc_N4Tm} shows the  classification accuracy as a function of temperature for this  $N_f=4$ network. Here,  the  low-temperature, low-$|m|$ configurations are those that exhibit EDWs. 
This network already classifies about half of the EDW configurations correctly, but still shows many wrong classifications of both low- and high-temperatures configurations, with an overall accuracy of 96\%.

\begin{figure}[t]
\includegraphics[width=0.9\columnwidth]{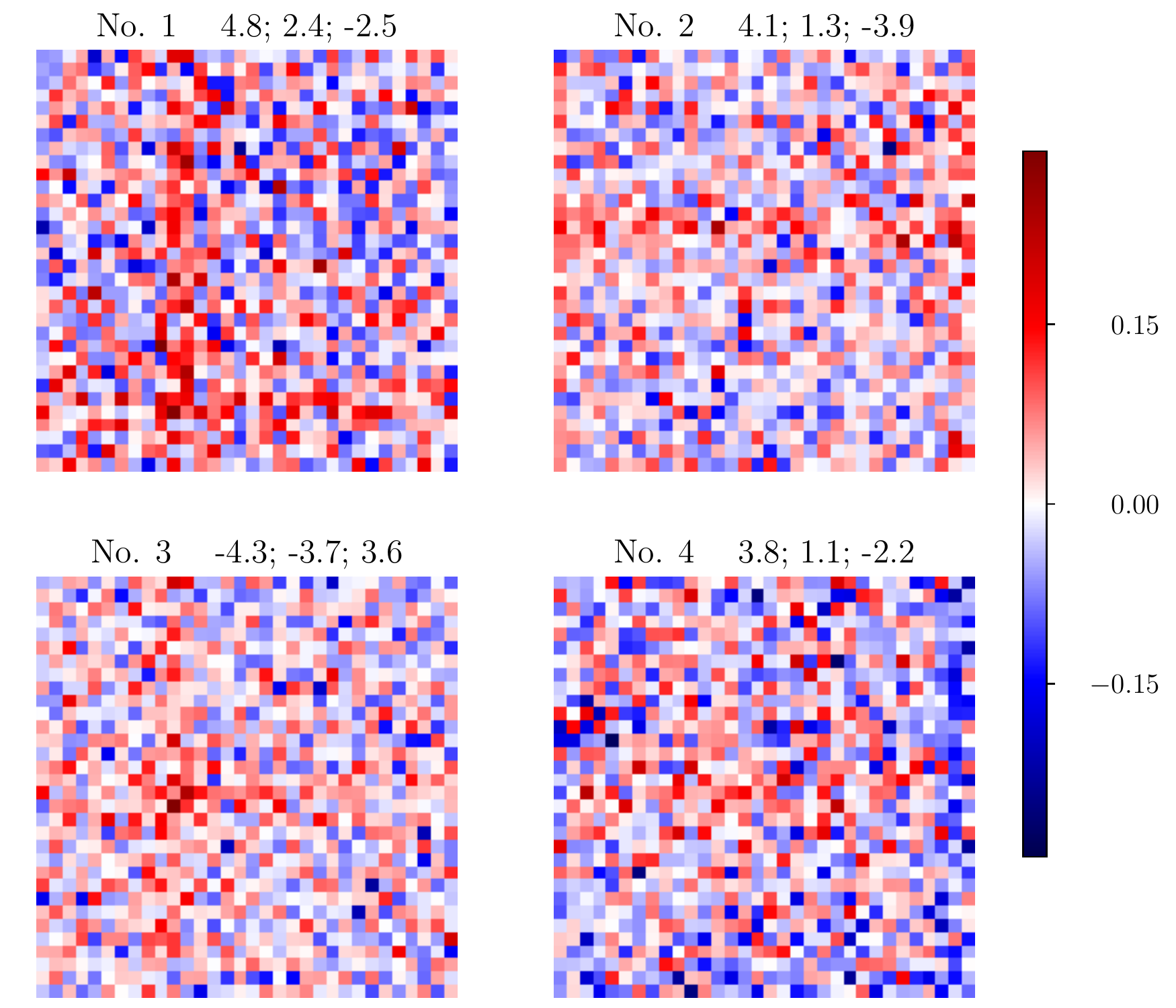}
\caption{(Color online). 
Weight matrices $w_{i,j}$ of the $N_f=4$ network for the Ising system with EDW configurations included. For each neuron (label by its index $i=1,...,4$), the weight matrix is shown as a two-dimensional array of $32\times 32$ values, corresponding to the layout of the  Ising model configurations. 
The numbers to the right of each neuron $i$ denote the local bias $b_i$, and the weights $w'_{1,i}$, $w'_{2,i}$ that connect this neuron to the   high- and low-temperature neuron of the  output layer (in this order). 
}
\label{fig_dw_fc_N4w}
\end{figure}

\begin{figure}[t]
\includegraphics[width=\columnwidth]{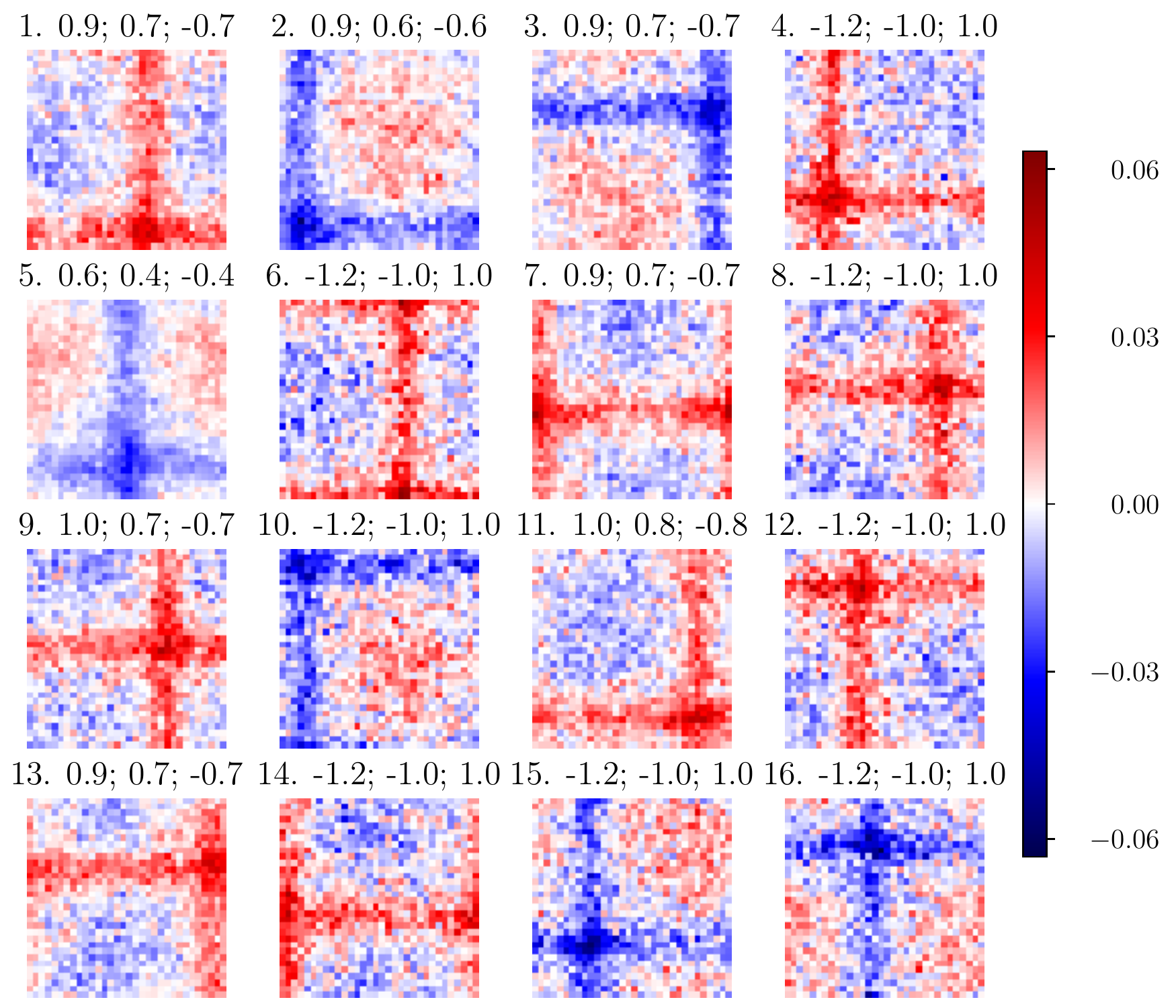}
\caption{(Color online). 
Weight matrices $w_{i,j}$ of the $N_f=16$ network for the Ising system with EDW configurations included. For each neuron (label by its index $i=1,..., 16$), the weight matrix is shown as a two-dimensional array of $32\times 32$ values, corresponding to the layout of the  Ising model configurations. 
The numbers to the right of each neuron $i$ denote the local bias $b_i$, and the weights $w'_{1,i}$, $w'_{2,i}$ that connect this neuron to the   high- and low-temperature neuron of the output layer (in this order). }
\label{fig_dw_fc_N16w}
\end{figure}

\begin{figure}[t]
\includegraphics[width=\columnwidth]{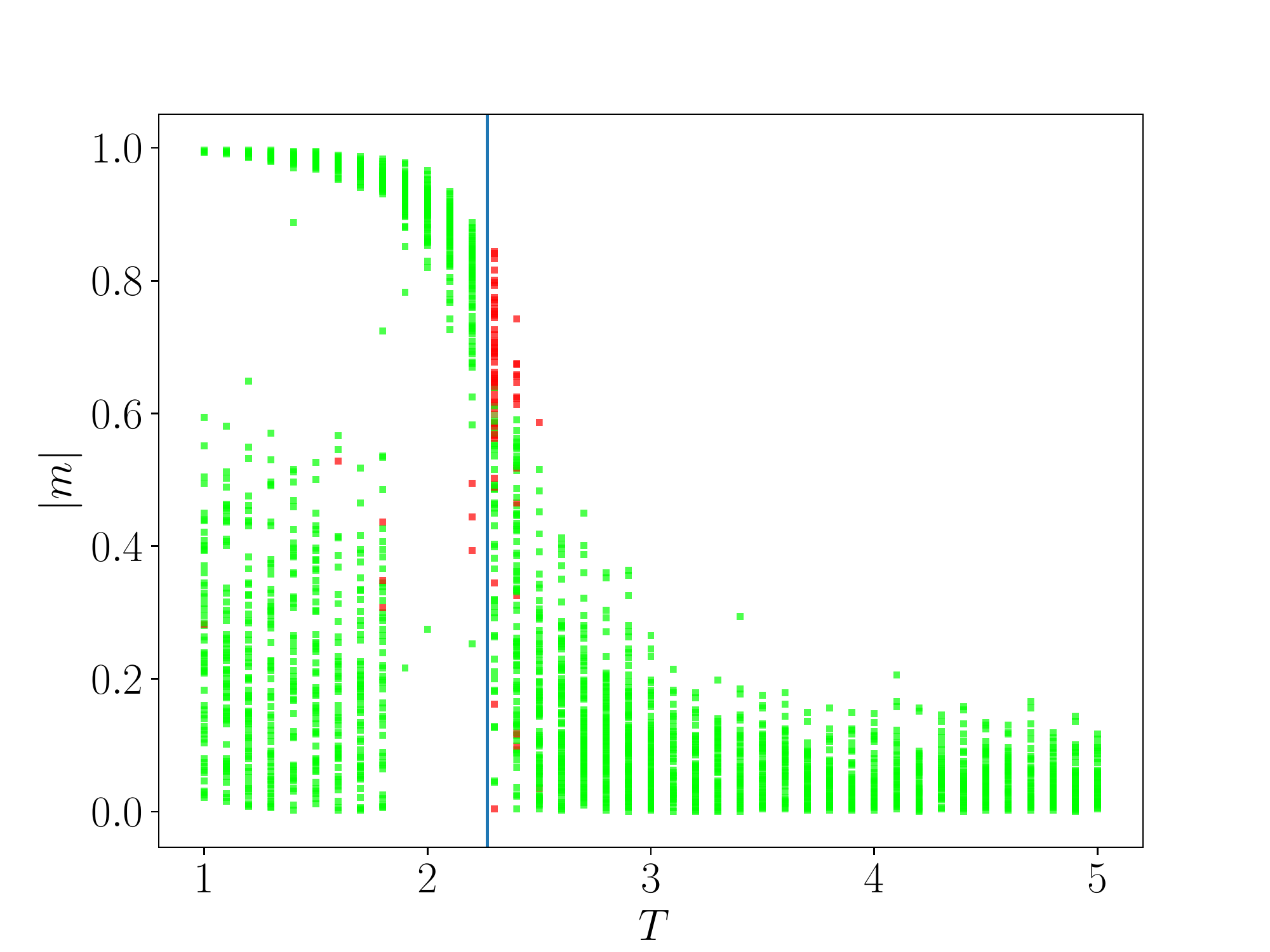}
\caption{(Color online).
Classification correctness  of the $N_f=16$ network for the Ising model with EDW configurations included, in the temperature $T$ vs.  magnetization $|m|$ plane of the input configurations. Green (red) dots indicate to a corrects (wrong) classifcations. The vertical line denotes the exact  transition temperature $T_c$. }
\label{fig_dw_fc_N16Tm}
\end{figure}

\begin{figure}[t]
\includegraphics[width=0.9\columnwidth]{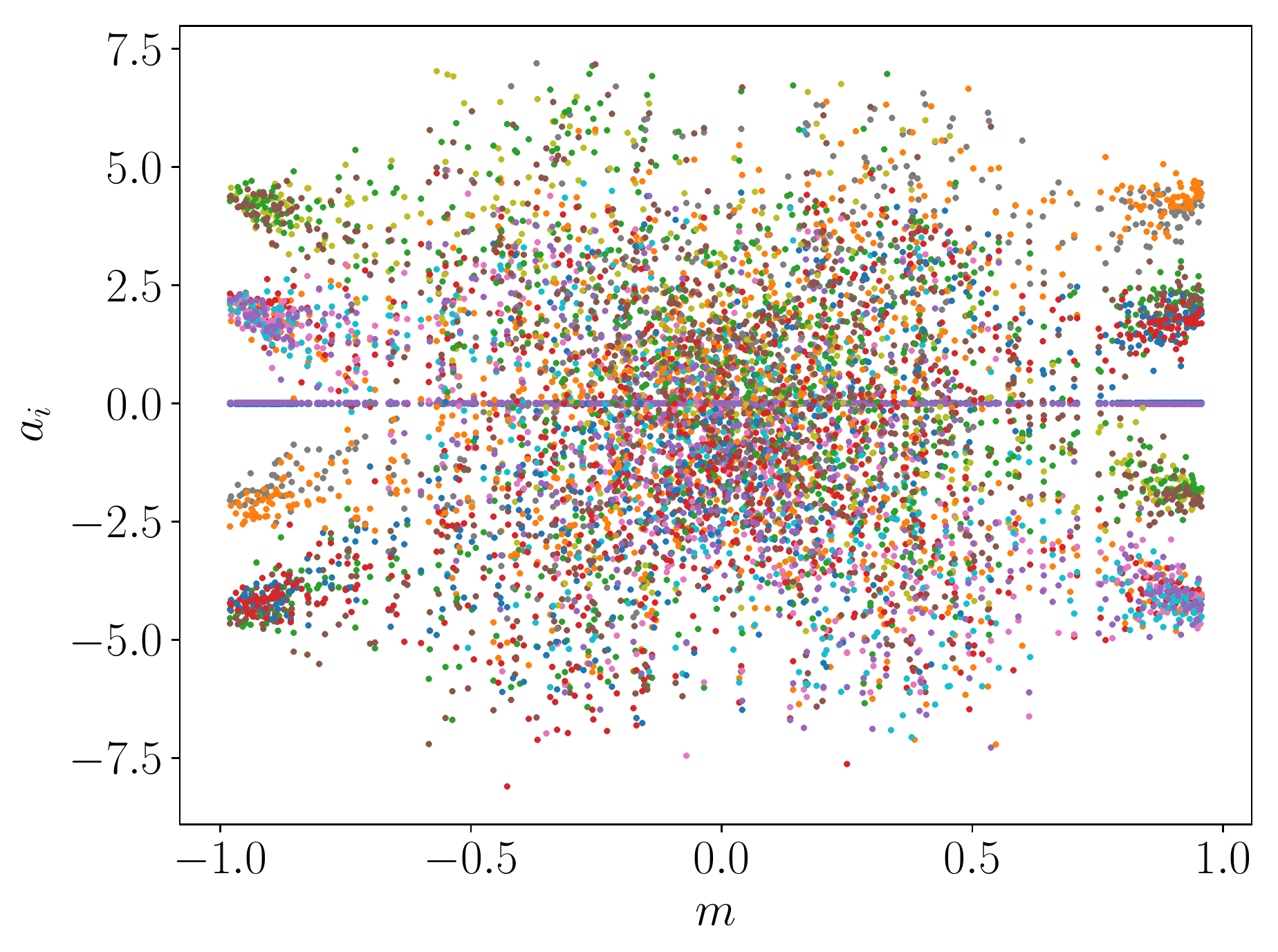}
\caption{(Color online). 
Hidden layer neuron activations $a_i$, $i=1,...,16$ for the $N_f=16$ network as a function of the magnetization $m$ of the input configuration.
Different colors denote the various neurons. 
}
\label{fig_dw_fc_N16am}
\end{figure}

\begin{figure}[t]
\includegraphics[width=0.9\columnwidth]{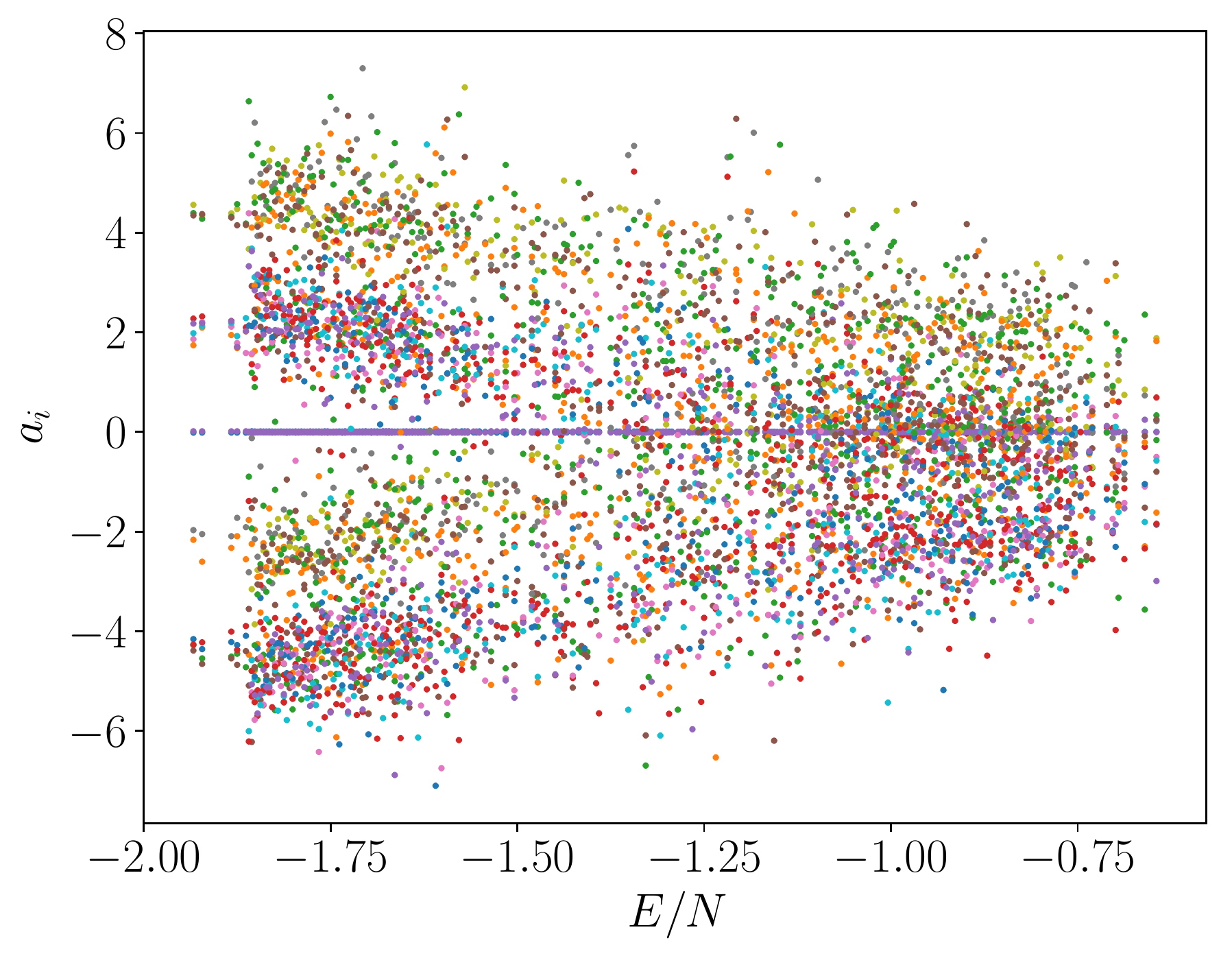}
\caption{(Color online). 
Hidden layer neuron activations $a_i$, $i=1,...,16$ for the $N_f=16$ network as a function of the energy $E$ of the input configuration.
Different colors denote the various neurons. 
}
\label{fig_dw_fc_N16aE}
\end{figure}

In order to rationalize the behavior of this network, we again
examine the weight matrices 
 $w_{i,j}$, which are shown in Fig.~\ref{fig_dw_fc_N4w}.  One can identify traces of strip-features in these weight matrices, most apparently for neuron No. 1.
 In order to exhibit these features more clearly, and to also improve on the  accuracy  of the network, we included more low-temperatures EDW configurations in the learning data set and 
 furthermore increased the number of hidden neurons to $N_f=16$. The final weight matrices for this network are  shown in Fig.~\ref{fig_dw_fc_N16w}. We identify  two striking features in these weight matrices: 
 (i)  each neuron exhibits a structure wherein a vertical and a horizontal stripe of strong polarization in the weight cross, and (ii) there is a similar number of such crossed-stripe neurons of both positive (red) and negative (blue) weights. These polarized domains in the weight matrices   reflect the predominantly vertical and horizontal orientation of  domains in the low-temperature EDW configurations. 
The neurons  are thus activated by regions in the input configuration that reside within an extended single domain. This allows the network to also classify the low-temperature EDW configurations correctly. This is seen explicitly in Fig.~\ref{fig_dw_fc_N16Tm}. The overall classification accuracy of this network is about 97\%.
%
%
Figure~\ref{fig_dw_fc_N16w} exhibits that the crossed-stripe structures for the different neurons are distributed broadly across the spatial domain. For a larger number of hidden neurons, a  more refined resolution of the various EDW configurations is of course possible, e.g.,  for a network with $N_f=64$, an overall classification accuracy of 98\% can thus be achieved. As the weight matrices show  crossed-stripe structures of both signs and at various positions, 
 the network may  be said to reflect  both the $Z_2$ symmetry of the Ising model  as well as its translational invariance. 
Examining  in Fig.~\ref{fig_dw_fc_N16am} the activations $a_i$ of the various neurons as a function of the magnetization $m$,
we find no clear overall relation between the hidden layer activations and the magnetization
apart from the regions of large magnetization $|m|$.
In  a plot of the activations vs. the energy $E$,  cf. Fig.~\ref{fig_dw_fc_N16aE}, 
we  also observe a rather broad overall distribution, which however exhibits  traces of a linear  relation between the activations and  the configurational energy. 
The network  apparently now performs the classification task based on a combined representation of the energy and the magnetization for the strongly polarized regime. \\

\begin{figure}[t]
\includegraphics[width=\columnwidth]{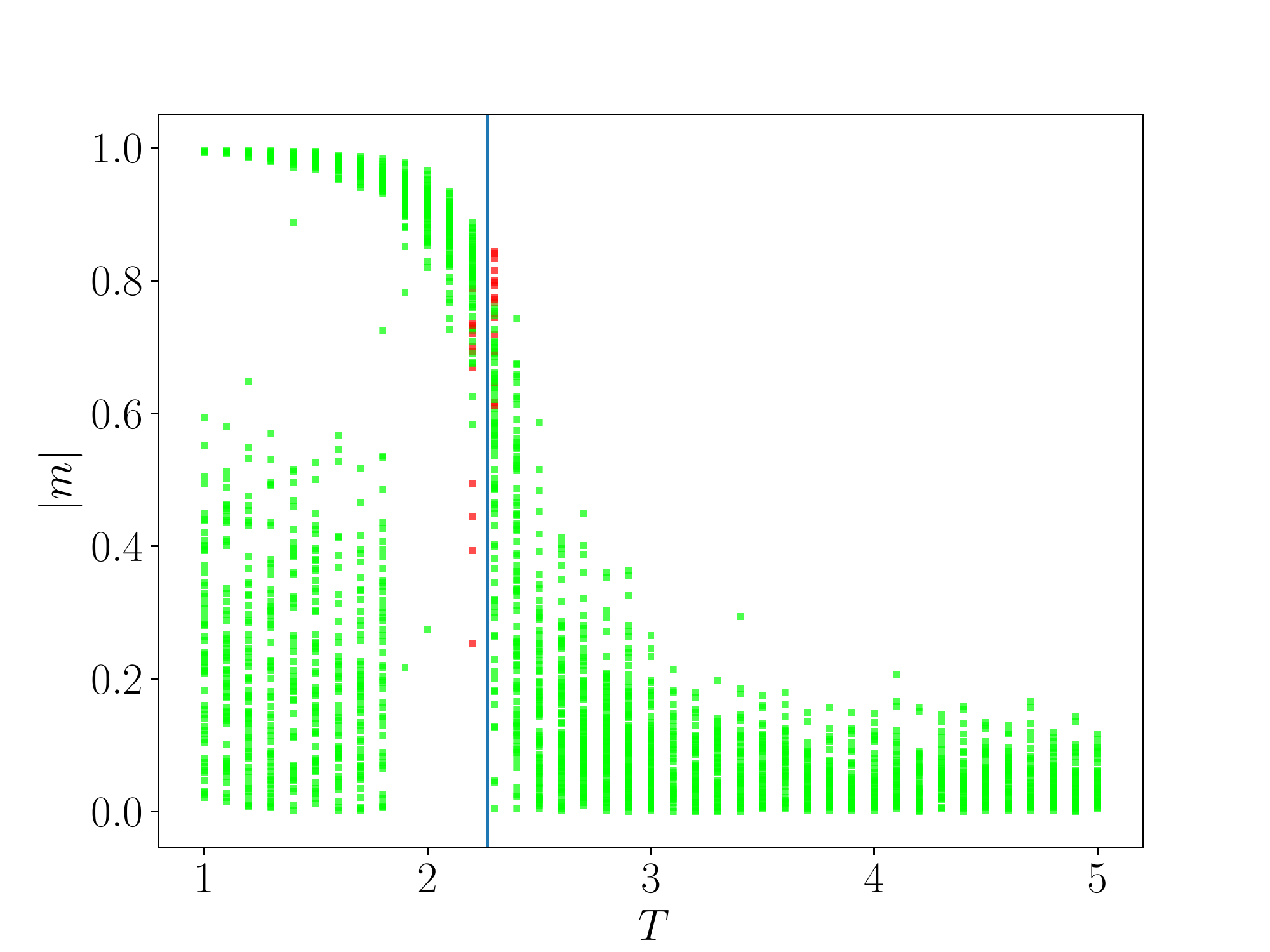}
\caption{(Color online).
Classification correctness  of the CNN with $N_k=8$  filter kernels of size $2\times 2$  for the Ising model with EDW configurations included, in the temperature $T$ vs.  magnetization $|m|$ plane of the input configurations. Green (red) dots indicate  correct (wrong) classifications. The vertical line denotes the exact  transition temperature $T_c$. 
}
\label{fig_dw_cnn_Tm}
\end{figure}

Based on  the above analysis, we expect that  already a shallow CNN will be able to  perform the classification task quite efficiently, since after training, its filters can 
readily identify the local contributions to the magnetization as well as  the local configurational energy. In the following, we demonstrate, that this is indeed the case. 
For this purpose, we trained a CNN with $N_k=8$ filter kernels in a single two-dimensional convolutional layer, similar to the CNN layout of Ref.~\onlinecite{Carrasquilla17}. Each filter kernel $K^{(k)}$, $k=1,...,N_k$ is a matrix of fixed size 
$2\times 2$ and uses the ReLU activation function. The output of this convolutional  layer is  then passed on to a  fully connected hidden layer with $N_f=16$ neurons. For each position $j$ of the $k$-th filter across the input layer, 
a weight matrix $w^{(k)}_{i,j}$ connects the output $z^{(k)}_j$ of this filter to the $i$-th neuron of the  fully connected hidden layer. In the case of the CNN,
the activation of the $i$-th neuron is thus given by $a_i=\sum_{k,j} w^{(k)}_{i,j} z^{(k)}_j +b_i$,  with  the local bias $b_i$, and using ReLU activation. 
Finally, each neuron $i$ of the  fully connected layer is  connected through weights $w'_{1,i}$ and $w'_{2,i}$
to the output layer with 2 neurons using softmax activation, as above. 
After training on the previously used data set (i.e., including the low-$T$ EDW configurations), the CNN
exhibits a high  classification accuracy of 99\%, as seen in Fig.~\ref{fig_dw_cnn_Tm}, where misclassifications are now constrained to the close vicinity of the phase transition. 

\begin{figure}[t]
\includegraphics[width=0.9\columnwidth]{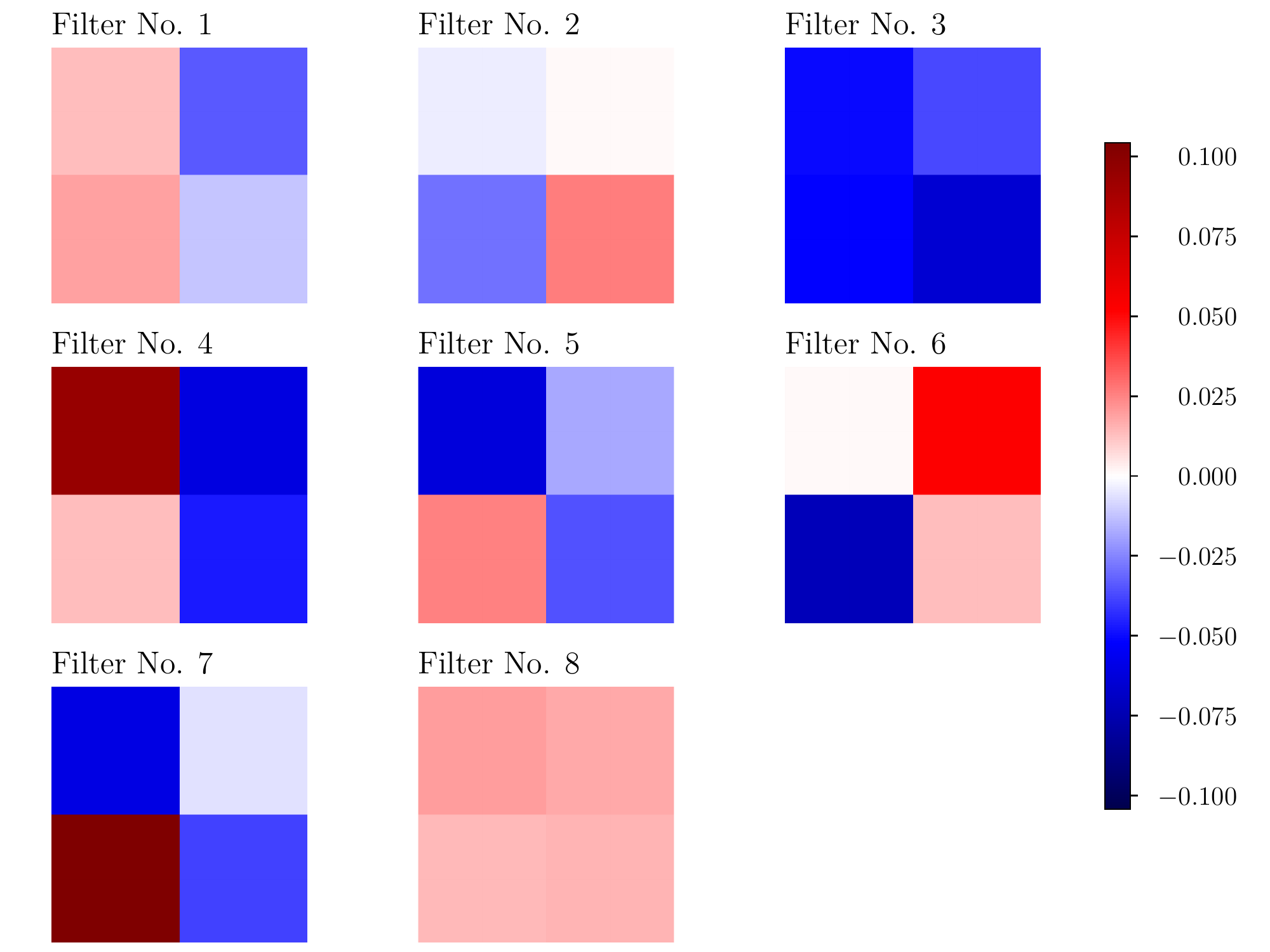}
\caption{(Color online).
Filter kernels $K^{(k)}$ of the $N_k=8$  CNN after learning the Ising model with EDW configurations included. 
}
\label{fig_dw_cnn_fk}
\end{figure}

In order to understand, how this CNN works, we examine directly the filter kernels, which are shown in Fig.~\ref{fig_dw_cnn_fk}. Some filters (in particular No.~3 and No.~8)  collect
a local average of the input values, where the different signs of the filter kernels reflect the $Z_2$ symmetry of the Ising model.  Most of the other filters (consider in particular No.~1, 2, 4, 6 and 7)  identify local domain walls in the input data, with  different orientations and signs. After the ReLU activation, only positive gradients in the corresponding direction are processed. 
For example,  filter No.~1 can identify a local domain wall oriented along the vertical direction.

\begin{figure}[t]
\includegraphics[width=0.9\columnwidth]{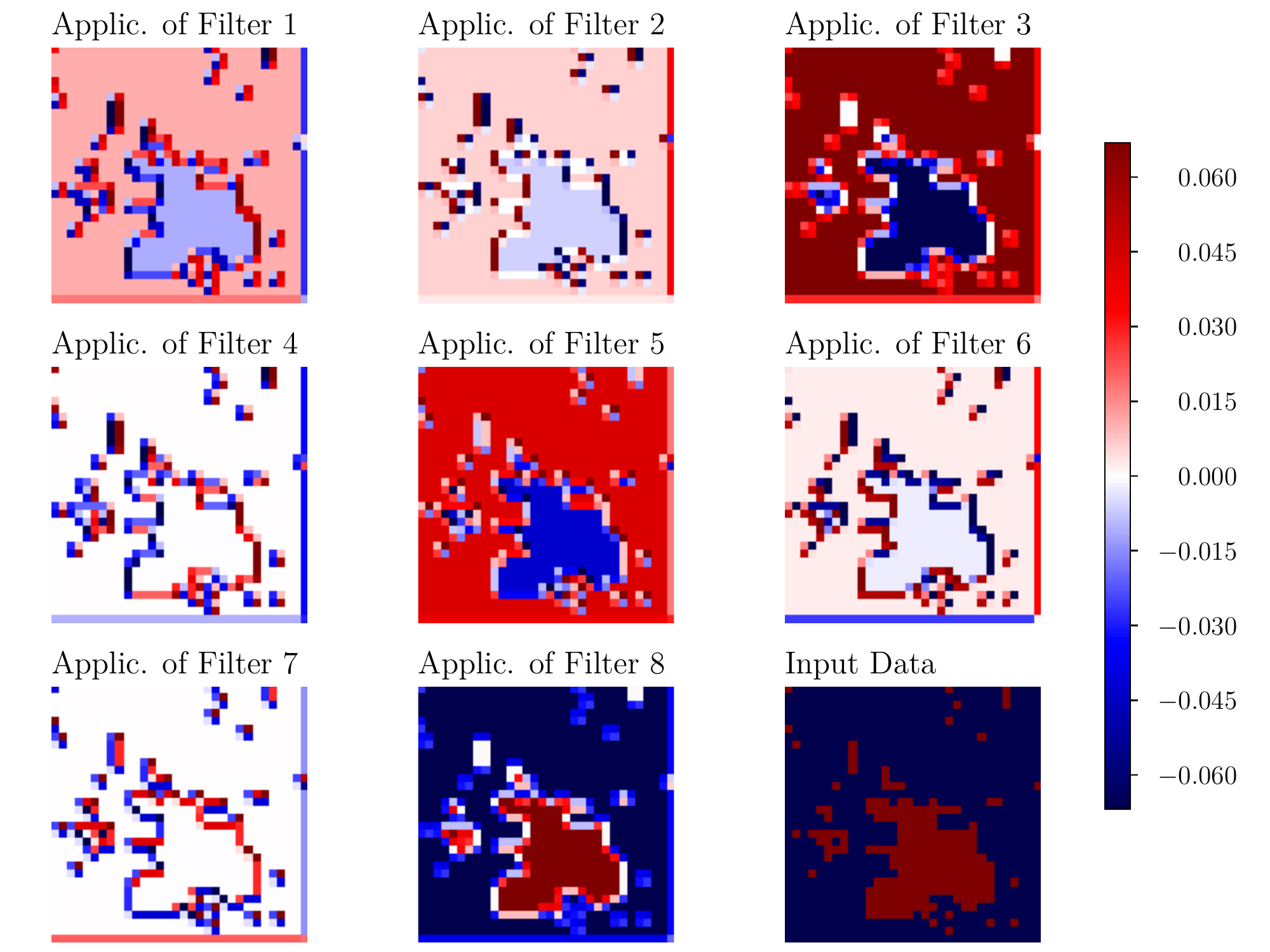}
\caption{(Color online).
Application of the $N_k=8$  filters of the CNN to the input configuration shown in the bottom right panel prior to the application of the ReLU function.  After application of the ReLU activation function, all blue regions (negative activation) will be set to zero. 
}
\label{fig_dw_cnn_sa}
\end{figure}

The effects of the various filters on the input data can be seen explicitly  by examining the  application of each filter to a given input configuration. 
These are shown in Fig.~\ref{fig_dw_cnn_sa}, for the specific input configuration that is shown in the bottom right panel. One observes that the filters No.~3, 5, and 8  essentially  propagate the averaged local magnetization of the input configuration, whereas the other filters specifically locate local domain boundaries. This is equivalent to calculating the local energy, depending on whether two neighboring spins are parallel  or not. 
Upon summation, the network thus estimates the configurational energy. 
In addition to the energy, the network  however also uses an  estimate of the overall magnetization, upon the summation of the output from the other filters.

\begin{figure}[t]
\includegraphics[width=0.9\columnwidth]{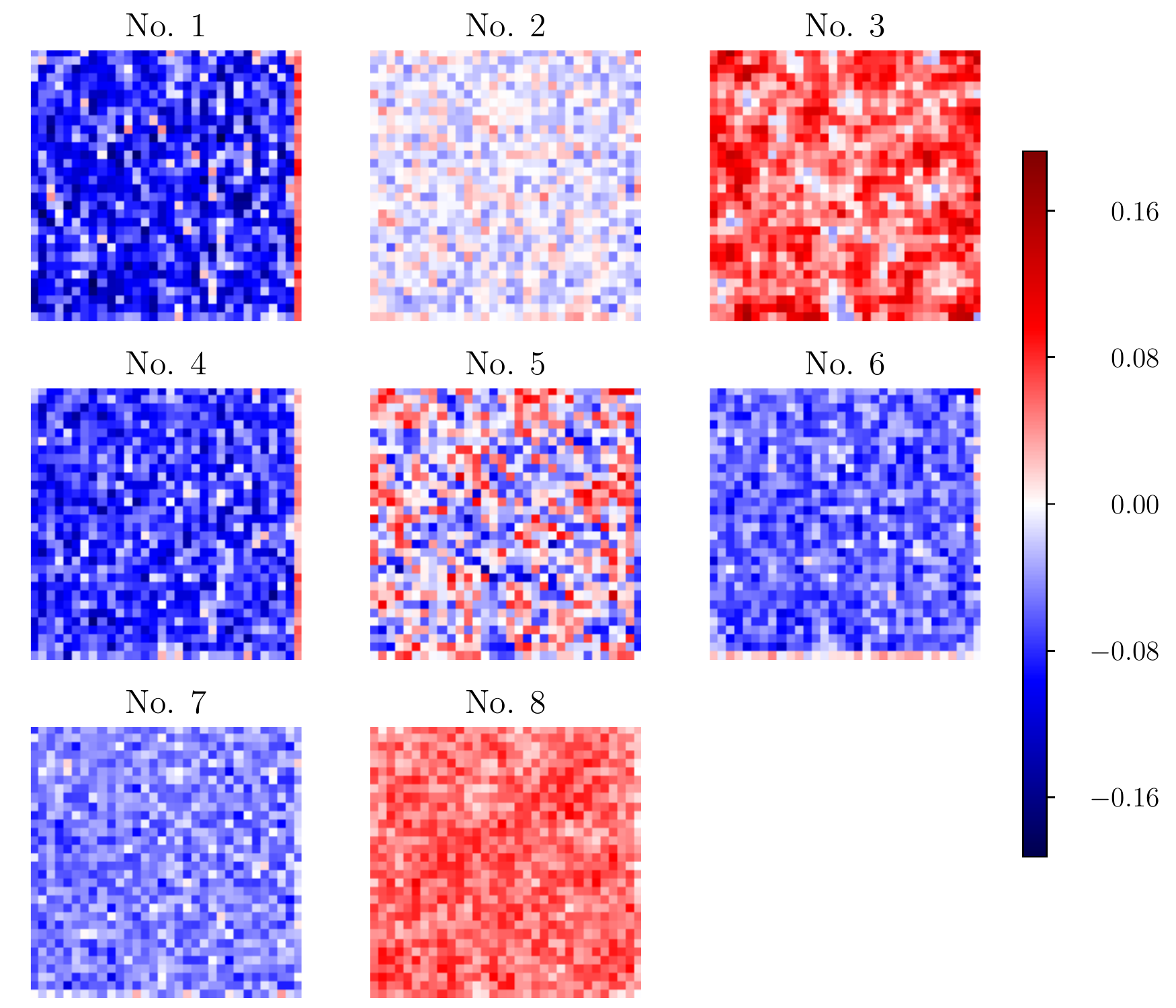}
\caption{(Color online).
Weight matrices $w^{(k)}_{i,j}$ of one of the neurons $i$ from the fully connected layer for each of the $N_k=8$ filters of the CNN after learning the  Ising model with extended domain wall configurations included.
For each filter kernel (label by its index $k=1,...,8$), the weight matrix is shown as a two-dimensional array of $32\times 32$ values, corresponding to the layout of the  Ising model configurations. 
}
\label{fig_dw_cnn_10}
\end{figure}

The way  the network gathers all this information together can be extracted from the example shown in Fig.~\ref{fig_dw_cnn_10}. Here, we display the  weight matrixes of one of the neurons from the fully connected layer. Each matrix connects this specific neuron to one of the $N_k=8$ filters. The element-wise multiplications are then summed up to form the activation of this neuron (for this specific neuron, the local basis turned out to be zero). Furthermore,  with respect to its connections to the output layer neurons, this neuron is  low-temperature activating (more specifically, its contributions to the output layer are 0.3 (0.7) for the high (low) temperature active output-layer  neuron). 
This fact can also be deduced from the following two features in  Fig.~\ref{fig_dw_cnn_10}: (i) the weight matrixes that relate to the domain-boundary filters contribute negatively, and rather uniformly, to the summation, while (ii) those  related to the magnetization contribute positively. Therefore, for a magnetization that is non-zero locally, and  a low amount of domain-boundaries in the input configuration, this neuron activates and contributes to the prediction of the low-temperature phase. The other neurons of this fully connected layer proceed similarly;  in particular, the neurons that  activate the high-temperature output neuron have weights of opposite signs (and are more noisy).

\begin{figure}[t]
\includegraphics[width=\columnwidth]{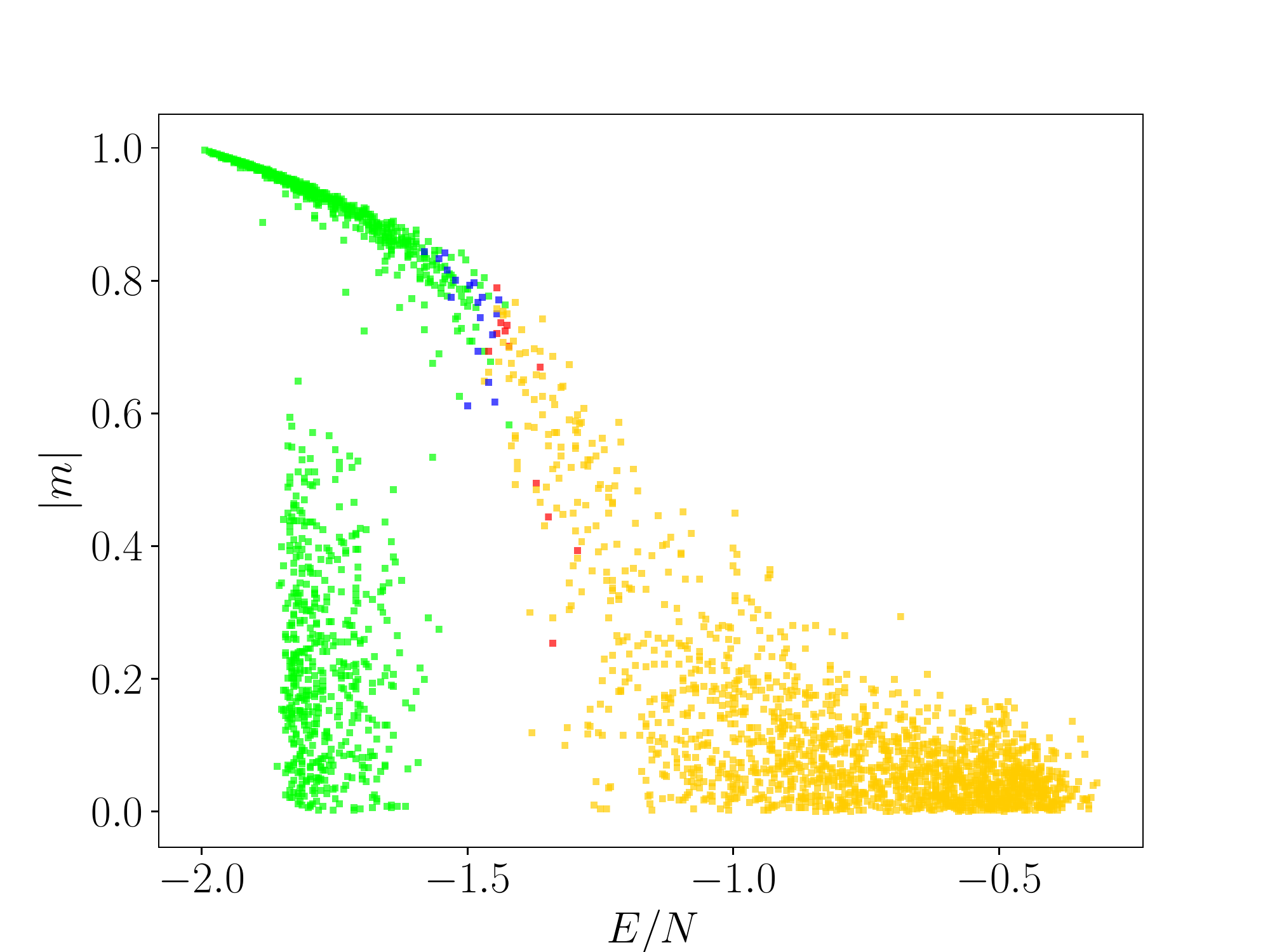}
\caption{(Color online).
Classification correctness  of the CNN with $N_k=8$  filter kernels of size $2\times 2$  for the Ising model with EDW configurations included, in the energy $E/N$ vs.  magnetization $|m|$ plane of the input configurations. 
Green (red) dots indicate correct (wrong) classifications as low temperature configurations, and orange (blue) dots  correct (wrong) classifications as high temperature configurations. 
}
\label{fig_dw_cnn_em}
\end{figure}

In summary, the CNN  uses threshold parameters  that essentially consist of the energy and the magnetization, based on which the final classification is  made. The fact that the energy plays an important role for the classification process of this CNN can also be seen in Fig.~\ref{fig_dw_cnn_em}, which shows the classification correctness in the energy vs. magnetization plane:  
a given configuration is seen to be  classified (correctly or wrongly)  to the low- or high-temperature phase based on a  dividing line at $E/N\approx 1.4$.
We expect  energy estimates to be effective for classification tasks of neural networks also in other cases, given that a simple estimate of its value is accessible by  filters that probe local gradients in the input values along  different lattice directions. A further example will be considered in the following section. 

\section{Supervised learning  the XY model}
\label{Sec:XY}
Another basic model of statistical physics that exhibits a finite temperature  transition between two distinct phases is the classical XY model, which is described by the Hamiltonian
\begin{equation}
H_{XY}=-J \sum_{\langle j,j' \rangle} \cos(\phi_j-\phi_{j'}),
\end{equation}
where the angles are constrained to the finite interval $\phi_j\in[0,2\pi)$.  We again consider an $N=L\times L$ sites square lattice geometry with periodic boundary conditions and fix units to $J=1$.  In the thermodynamic limit, this model exhibits a
Kosterlitz-Thouless transition at a transition temperature of $T_\mathrm{KT}=0.893$, which is driven by the proliferation of vortices, topological point-defects in the spin configuration~\cite{Kosterlitz73}. Upon lowering  the temperature,  these vortices  confine into vortex anti-vortex pairs, and below $T_\mathrm{KT}$
the system shows  an algebraic decay of the  spin-spin correlations. In accord with the Mermin-Wagner theorem~\cite{Mermin66},  long-range order with a finite order-parameter is constrained to the zero-temperature limit. 
The high-temperature phase  instead shows  an exponential spatial decay of the spin-spin correlations, with a correlation length that diverges exponentially upon approaching $T_\mathrm{KT}$.
The  temperature region just above $T_\mathrm{KT}$ is dominated by  an enhanced proliferation of entropy from the unbinding of the vortex anti-vortex pairs. This  results in a (nonuniversal) peak in the specific-heat at a distinct temperature of $T_\mathrm{max}\approx 1.1$, slightly above the 
actual phase transition  at $T_{KT}$, whereas the specific heat $C$ does not exhibit a peak at $T_\mathrm{KT}$~\cite{Chaikin00}. 
While the phase transition in the XY model is  driven by vortices, i.e., by  topological defects, 
it is  not clear,  to what extend their presence is also  useful  for the machine learning of  this phase transition -- in particular, if the spin configurations are directly taken as the input data, which would involve the least preprocessing. 
Indeed, on a finite lattice, 
the XY model essentially appears long ranged ordered well below $T_\mathrm{KT}$, e.g., the average value of the magnetization $|m|$, 
where $m=\frac{1}{N}\sum_j e^{i\phi_j}$ for the XY model, takes on finite values that reduce rather slowly with the system size~\cite{Chung99,Olsson95,Beach17} as compared to, e.g., the Ising model.
Therefore, a neural network may still simply learn to use the value of the finite-size magnetization to discriminate 
the ordered from the disordered regime, as was suggested recently in Ref.~\onlinecite{Beach17}.  The magnetization in finite-size samples has also been identified as a relevant quantity in other recent studies  of the XY model with unsupervised learning schemes, such as  in principle component analysis (PCA) or by variational autoencoders~\cite{Wetzel17b,Wang17,Cristoforetti17}.

\begin{figure}[t]
\includegraphics[width=0.9\columnwidth]{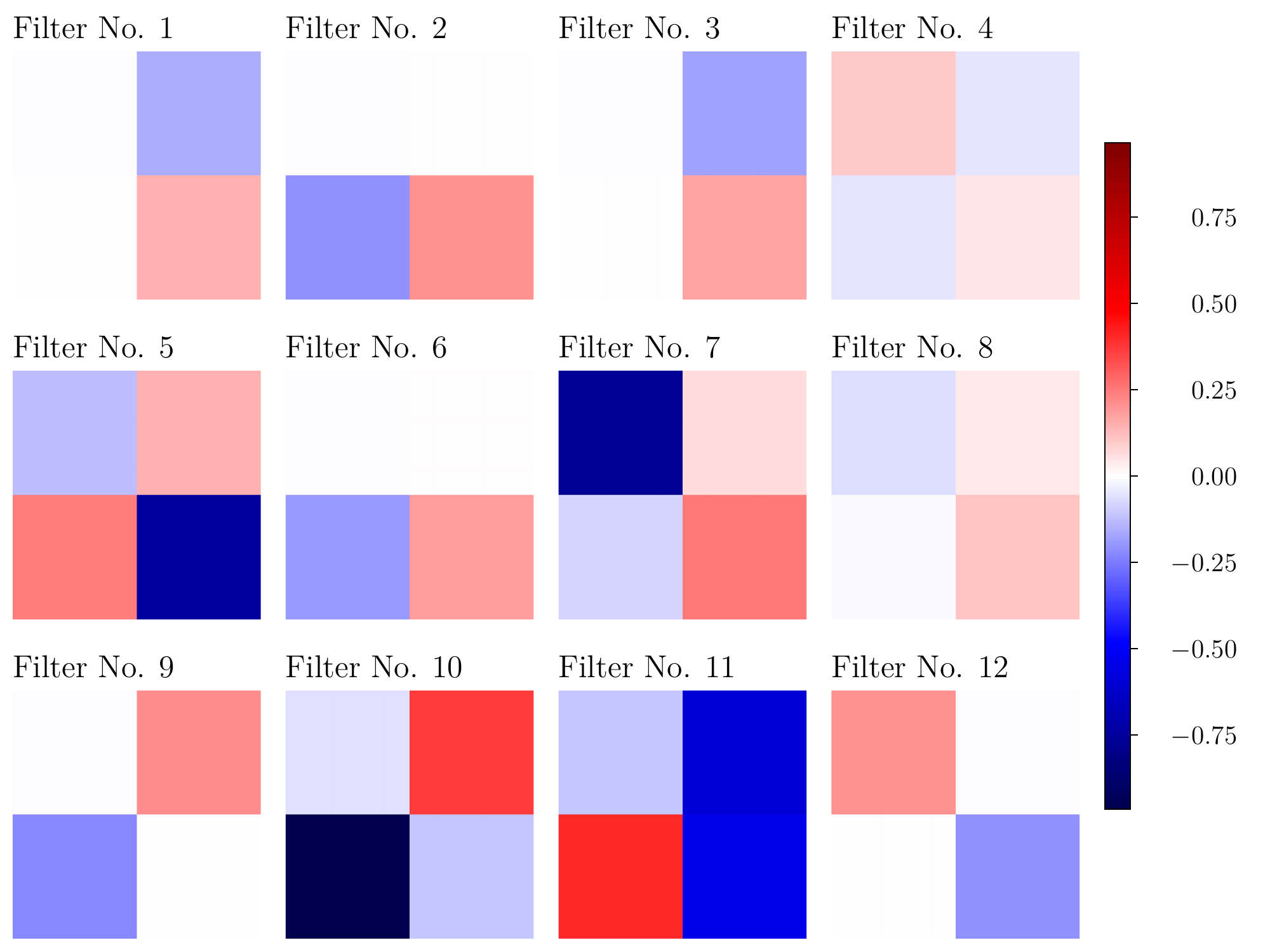}
\caption{(Color online). Filter kernels $K^{(k)}$ of the $N_k=12$  CNN after learning the XY model spin configurations.}
\label{fig_xy_k}
\end{figure}

To examine this issue in more detail for the case of a shallow CNN, we consider here again the CNN  with a single convolutional layer and  a kernel size of  $2 \times 2$, that was used in the previous section.
For the input signal $x_j$  of the $j$-th input neuron  in a given  configuration of the XY model, we rescaled the corresponding angle variable to $x_j=\phi_j/\pi$. 
In the following, we consider in particular a system with $L=32$, and a training data set over a temperature range between $T=0.2$ and $T=1.6$, with a spacing of $\Delta T=0.05$, obtained using the Wolff update scheme.

\begin{figure}[t]
\includegraphics[width=0.9\columnwidth]{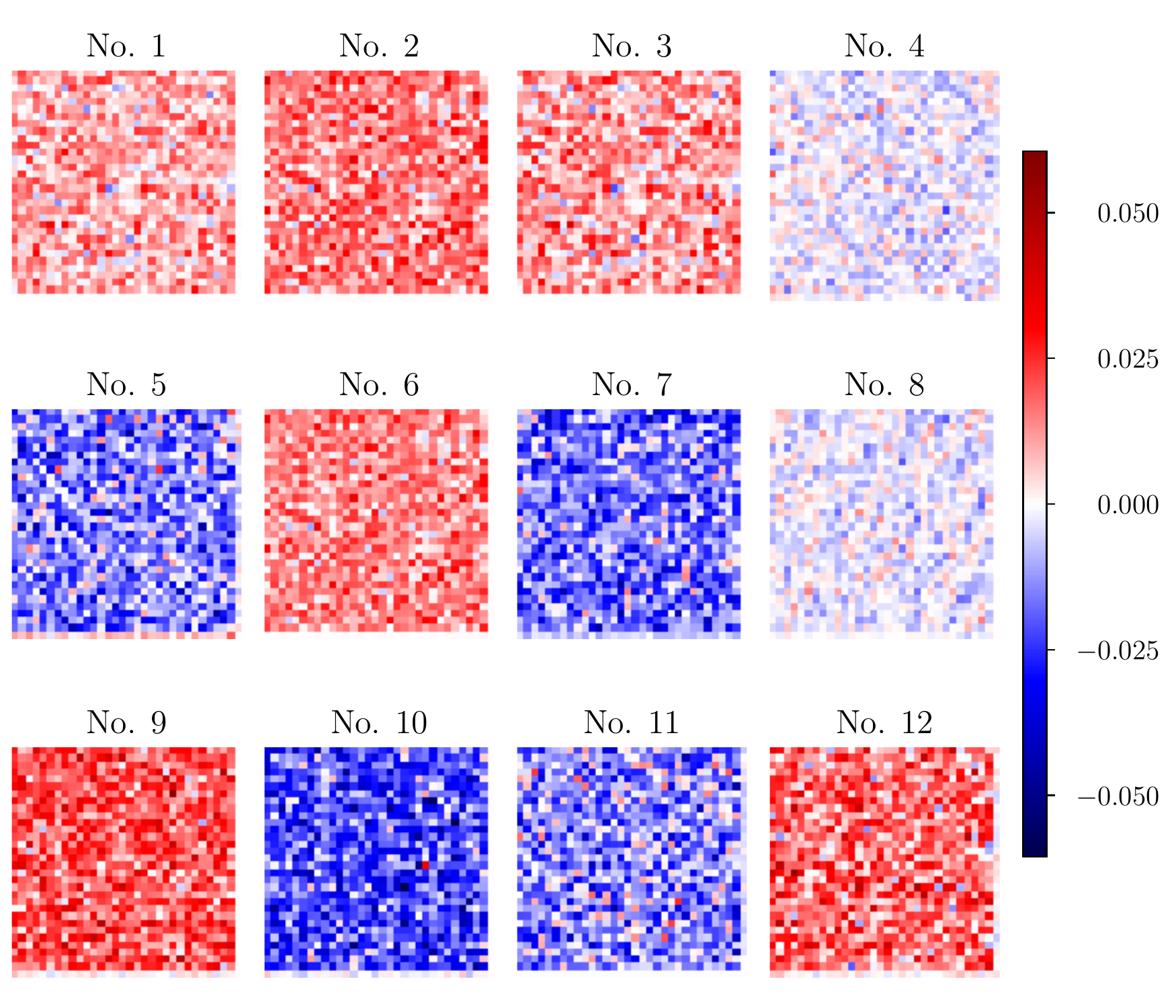}
\caption{(Color online). 
Weight matrices $w^{(k)}_{i,j}$ of one of the neurons $i$ from the fully connected layer for each of the $N_k=12$ filters of the CNN after learning the  XY model spin configurations.
For each filter kernel  (label by its index $k=1,...,12$), the weight matrix is shown as a two-dimensional array of $32 \times  32$ values, corresponding to the layout of the  XY model configurations. }
\label{fig_xy_w}
\end{figure}

The  final form of the kernels for a CNN with $N_k=12$  filters is shown Fig.~\ref{fig_xy_k}.
One can identify two major kernel classes for this network:  about half of the filters (No. 1, 2, 3, 6, 9, 12)  apparently identify local differences in the angles, along either vertical, horizontal or diagonal lattice directions. This  provides an estimate of the local gradients of the input configuration. We denote such filters as difference filters. However, due to the branch cut of the angular variables at the upper limit of the  finite interval $[0,2\pi)$, the sole presence of such difference filters would  lead to the false identification of large local angle differences: the network needs to learn that neighboring angle variables which differ slightly across $2\pi$, such as $2\pi-\epsilon$ and $2\pi+\epsilon$ (with $\epsilon\ll 1$), actually represent only a small local gradient. We find that  other filters  in Fig.~\ref{fig_xy_k} (No.~5, 7, 10, 11) apparently serve this purpose, and  thus denote them as correction filters. To illustrate this behavior,  Fig.~\ref{fig_xy_w}  shows as a representative example the weight matrices of one of the three neurons from the fully connected layer for each of the $N_k=12$ filters. We observe dominantly positive matrix elements for the difference filters, while the weight matrices 
that connect to the filters No.~5, 7, 10, 11 are dominantly negative, and thus counteract  the activation from the difference filters
 (the other two filters, No.~4 and 8, in addition to having small kernel values,  contribute to the activation of the fully connected layer through lower weight matrices, and are thus apparently less  important than the other filters). 
\begin{figure}[t]
\includegraphics[width=\columnwidth]{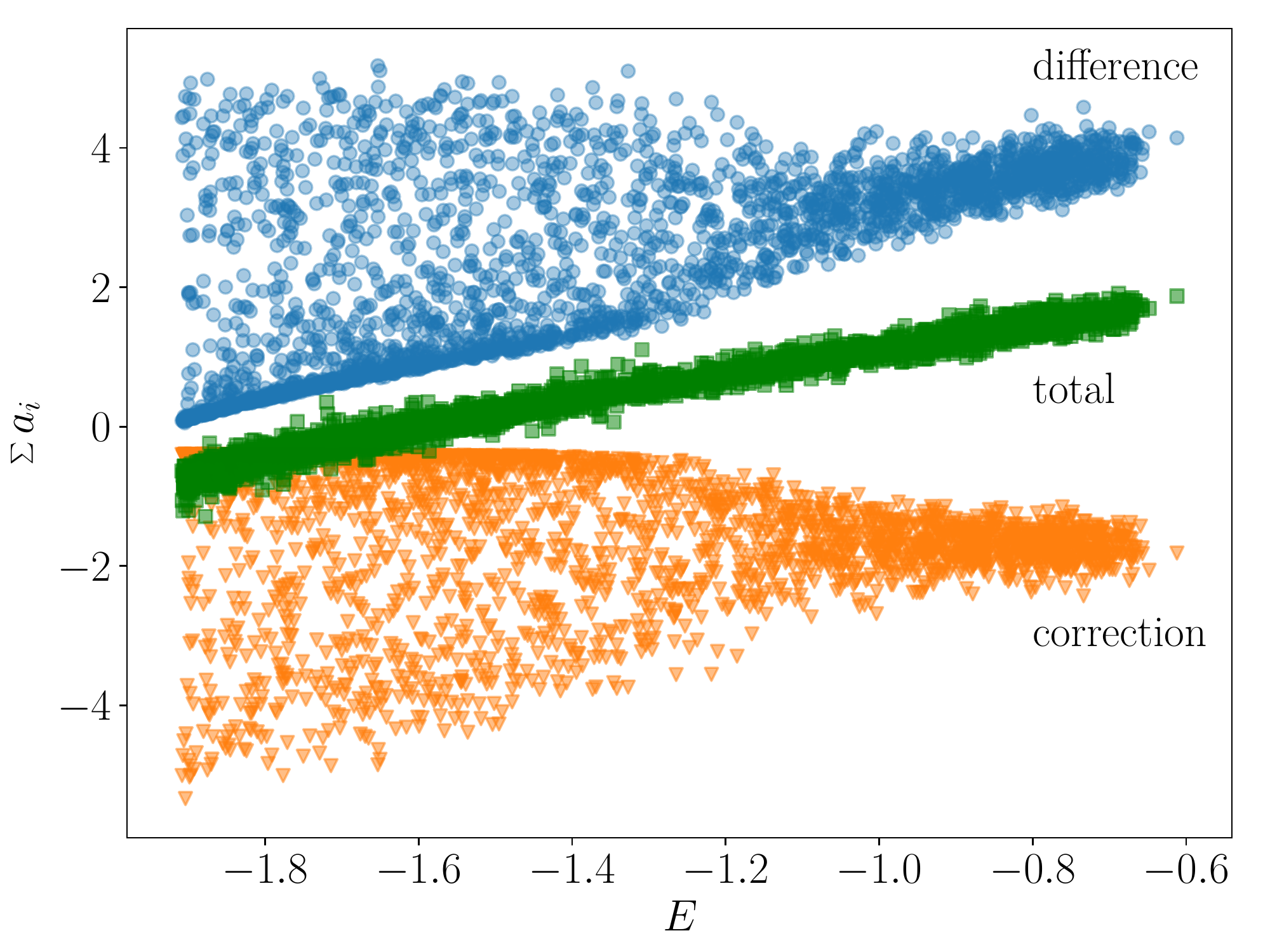}
\caption{(Color online). Total activation (squares, green), partial activation from the difference filters (circles, blue), and partial activation from the correction filters (triangles, orange)  as functions of the configurational energy $E/N$ for one of the three neurons  
of the $N_k=12$  CNN after learning the  XY  model configurations.  
}
\label{fig_xy_a}
\end{figure}

If one  plots  the contributions to the  activation of this specific neuron from  the difference filters  vs. the  energy of the input XY model configuration, one  obtains the positive data shown in  Fig.~\ref{fig_xy_a}.  Accordingly, the separate contributions to the activation from the correction filters result in negative values in Fig.~\ref{fig_xy_a}. Moreover, both of these separate contributions to the activation show a rather broad spread of values, in particular in the low-energy region. Remarkably however, upon summing the weighted contributions from {\it all} filters, cf. Fig.~\ref{fig_xy_a}, the resulting total activation of this neuron exhibits a narrow, essentially linear scaling with the configurational energy. Given that the rather broad spread seen in the  contributions to the activation from the difference filters is due to the  
presence of local angle differences across the branch cut,  these are thus corrected for by the correction filters (the total result is indeed very similar, if one sums over all filters except No. 4 and 8, which have low weights). 
This combined information is then processed further to the output layer in order to perform the final classification task. 
Further insights into the workings of the CNN  for the case of the XY model can  also be obtained by using it in the learning-by-confusion scheme of Ref.~\onlinecite{Nieuwenburg17}, which we consider in the next section.

\section{Confusion learning the XY model}\label{Sec:Confusion}

The learning-by-confusion scheme of Ref.~\onlinecite{Nieuwenburg17} tries to estimate the phase transition temperature $T_c$ from the classification performance of the neural network within a given temperature range that contains $T_c$. 
For this purpose, the classification performance of the network is monitored
as a function of a  guess value $T^*$ for the actual  transition temperature as follows:
For a given value of $T^*$  from the considered temperature window, each learning set configuration is labeled 
into a high- or low-temperature class, depending on whether its temperature $T$ is above or below $T^*$. 
Based on this labeling, one trains the neural network as in the supervised learning scheme. 
After training,  the test accuracy for a given value of $T^*$ is then given by the relative number of test configurations 
that are correctly classified by the neural network. 
Under the assumption that
the network is capable of learning an appropriate parameter that relates to the physics of the phase transition, one  expects that the test accuracy of the
classification procedure exhibits a local maximum at a value of $T^*$ close to the true $T_c$. 
This is so, because for $T^*$ equal to $T_c$, the network experiences the least confusion in the behavior of the physical quantity and the class assignment based on $T^*$.
Furthermore, for values of $T^*$ near the end of the considered temperature range,
the network is being trained and tested on essentially one class only,  so that  a high test accuracy will result. Hence, as a function of
$T^*$, one expects a w-shape  to result in the test accuracy,  
thereby providing an estimate of the actual transition temperature $T_c$ (up to  finite-size effects)~\cite{Nieuwenburg17}.
As we will show in the following, one can employ this  scheme also as a diagnostic tool for of the underlying neural network design. 

\begin{figure}[t]
\includegraphics[width=1.1\columnwidth]{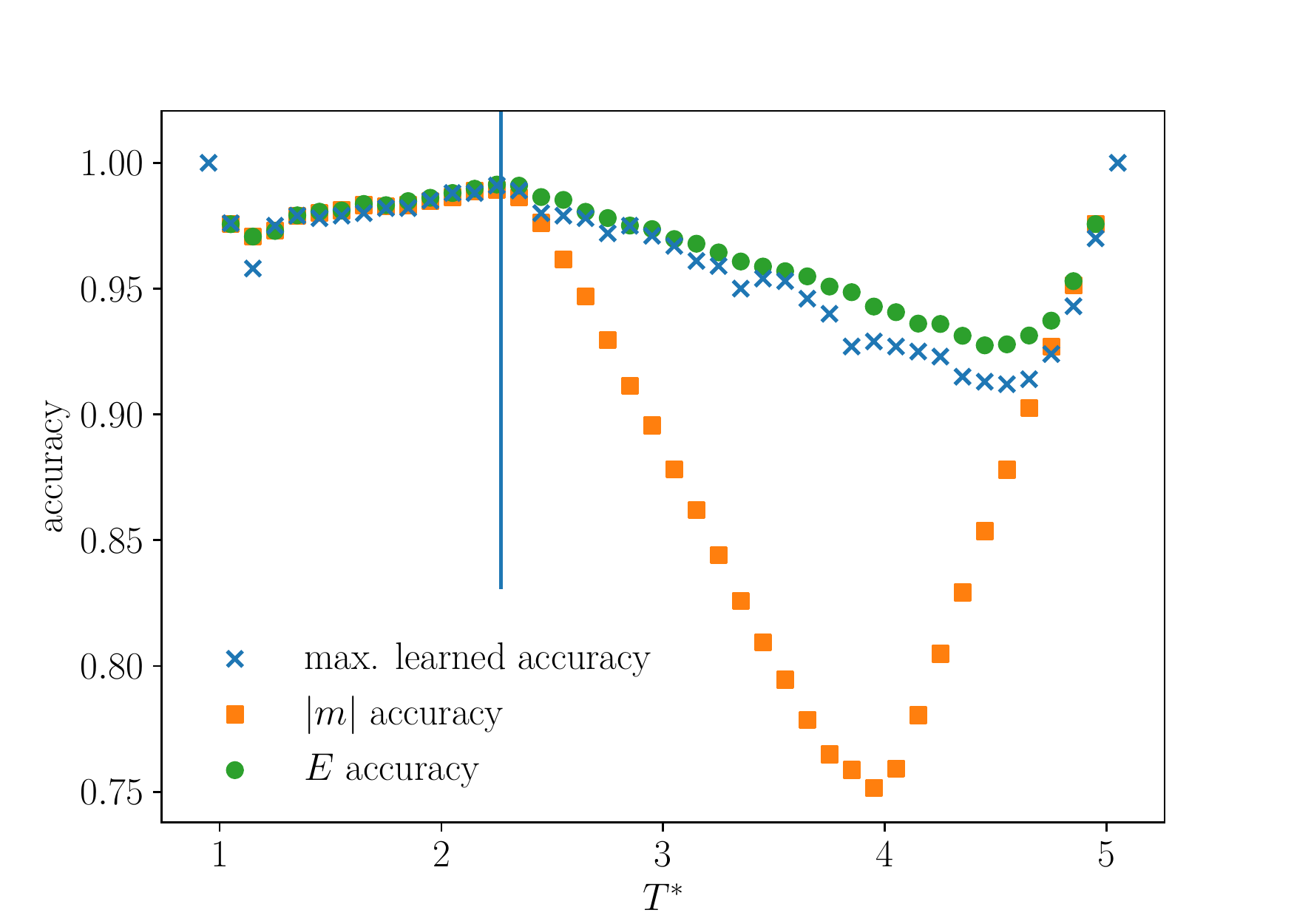}
\caption{(Color online). Test accuracy of the  learning-by-confusion scheme for the Ising model on a $L=32$ lattice without low-temperature EDW configurations. Also shown are the accuracies from the
threshold-value classification based on the magnetization $|m|$ and the configurational energy $E$. 
The vertical line denotes the exact  transition temperature $T_c$. 
}
\label{fig_cs_i} 
\end{figure}

\begin{figure}[t]
\includegraphics[width=1.1\columnwidth]{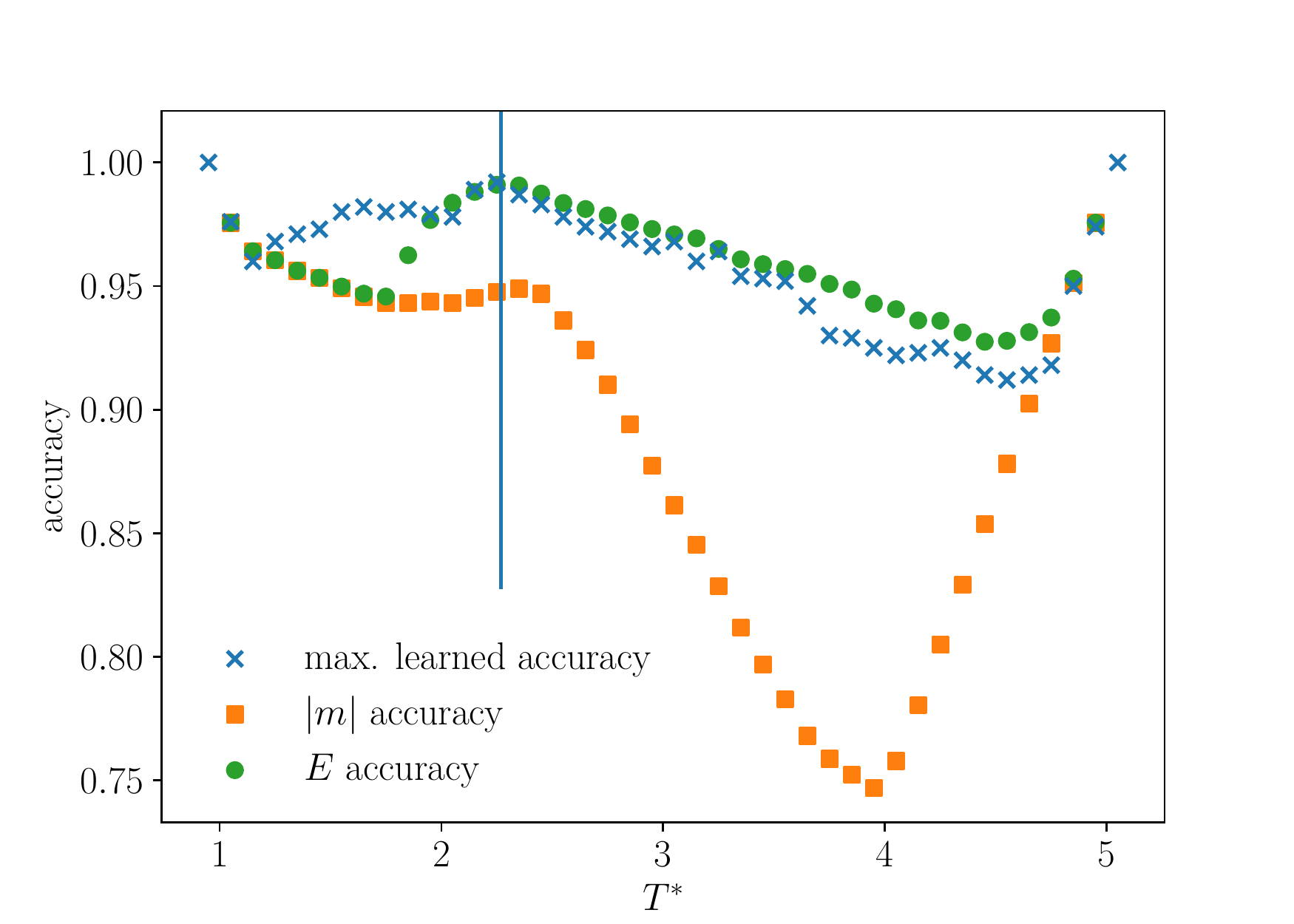}
\caption{(Color online). Test accuracy of the  learning-by-confusion scheme for the Ising model on a $L=32$ lattice in the presence of low-temperature EDW configurations. Also shown are the accuracies from the
threshold-value classification based on the magnetization $|m|$ and the configurational energy $E$. 
The vertical line denotes the exact  transition temperature $T_c$. 
}
\label{fig_cs_id} 
\end{figure}

Before applying for this purpose the confusion scheme  to the XY model, it will be useful to  reconsider  the case of  the Ising model~\cite{Nieuwenburg17}.
The resulting w-shape of the maximum achieved classification accuracy in the confusion scheme for  the CNN from Sec.~\ref{Sec:Domainwalls} as a function of  $T^*$ is shown in  Fig.~\ref{fig_cs_i}. Here, we first consider the case
that   no EDW configurations are included in both the learning and  the test configurations. 
In this figure, we also compare the  test accuracy of the CNN to 
a  simple threshold-value classification, based on specific physical quantities. This is shown in Fig.~\ref{fig_cs_i}  for two cases: the magnetization $|m|$ and the configurational energy $E$.

Both curves were obtained as follows: Consider a physical quantity $A$ (such as the energy $E$) that within the considered temperature range increases with temperature $T$ (if A  decreases with increasing $T$,  consider $-A$ instead)
and chose a threshold-value $A^*$ of $A$ for a given value of $T^*$, such as, e.g.,  the mean value $\langle A \rangle_{T=T^*}$ of $A$ at $T=T^*$. 
In the threshold-value classification, the phase assigned to a test configuration is then based on whether its value for $A$ is larger or lower than $A^*$, so that 
the test accuracy  equals the relative number of
sample configurations for which the differences $(A-A^*)$ and $(T-T^*)$ have the same sign.
Plotting this number as a function of $T^*$  provides the  threshold-value classification accuracy based on the considered observable $A$.
In practice, we observed that for a given value of  $T^*$, the accuracy of this classification can be  increased   
by optimizing the threshold-value $A^*$ in the vicinity of $\langle A \rangle_{T=T^*}$, e.g., by an iterative procedure. 
If the neural network would base its classification directly on a physical quantity $A$,  one would thus also  expect  its test accuracy in the learning-by-confusion scheme to follow 
the accuracy of the threshold-value classification based on $A$. 

For the Ising model, the test accuracy of the confusion scheme
in Fig.~\ref{fig_cs_i} actually rather closely
traces the accuracy of the threshold-value classification  based on  the energy $E$ over the full temperature range. This suggests that indeed the CNN uses an estimate of the configurational energy to 
perform the classification task, such that it may be said to have learned the energy. 
Moreover, the threshold-value classification  based on  the energy $E$  is found to be more accurate than based on the magnetization $|m|$ for values of $T^*$ above the critical temperature, while in the low-temperature regime they perform similarly well. This is due to the fact that above the transition temperature
the magnetization $|m|$ exhibits a much weaker temperature dependence than the energy $E$, so that the later can serve better as a threshold-value in this temperature region. 
If we   repeat this procedure for the Ising model with EDW configurations contained at low temperatures, we obtain the results shown in  Fig.~\ref{fig_cs_id}. 
While the CNN still shows a similarly high overall performance as in the absence of  EDW configurations, we  find that 
the  threshold-value classifications  based on  the energy $E$ and $|m|$ now fall below the CNN accuracy within the low-$T^*$ region. This 
is due to the fact that the CNN can 
correctly identify the EDW configurations, while these configurations 
have an increased energy $E$ due to the domain walls, and a corresponding low value of $|m|$, as discussed in Sec.~\ref{Sec:Domainwalls}.

\begin{figure}[t]
\includegraphics[width=1.1\columnwidth]{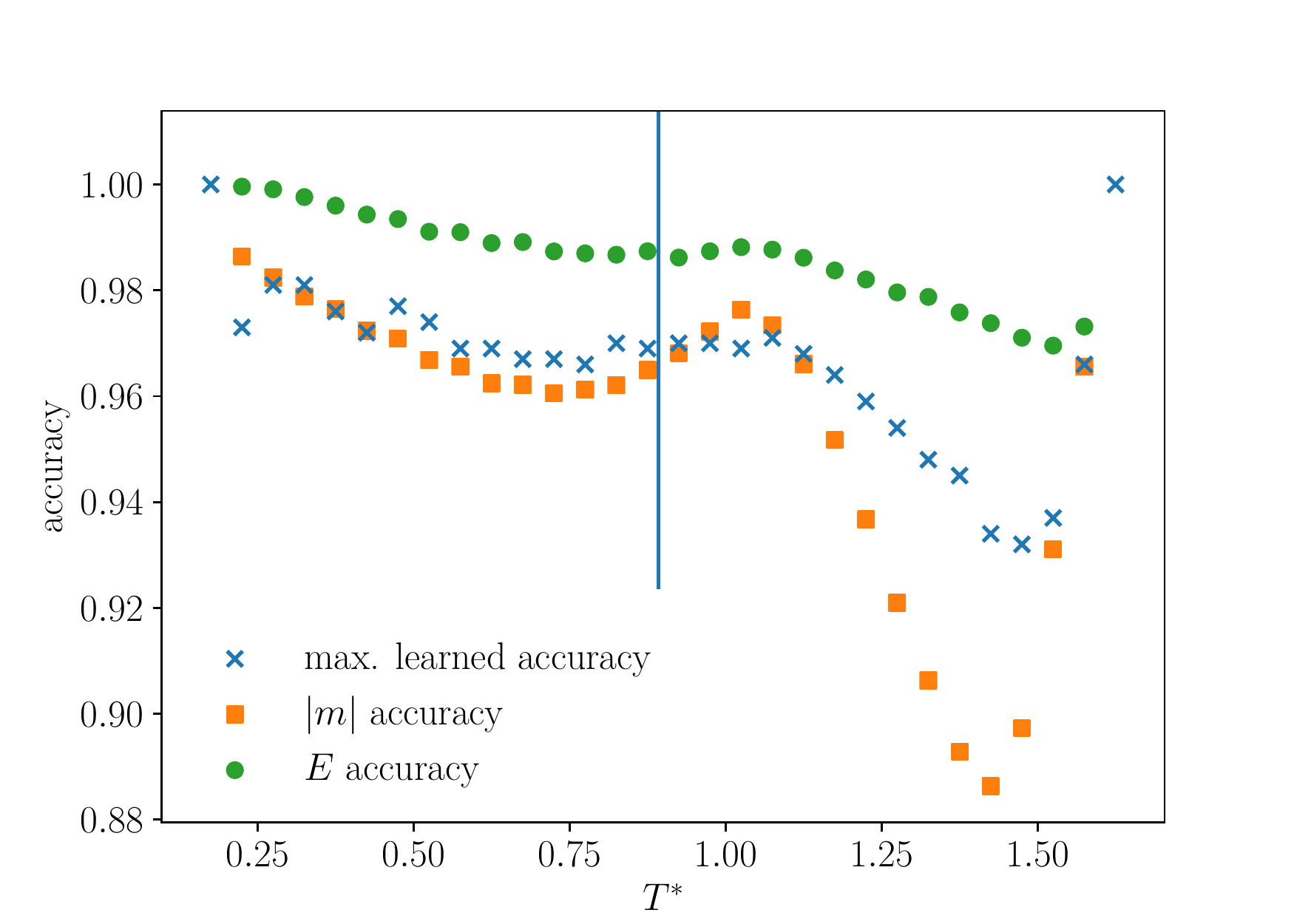}
\caption{(Color online). Test accuracy of the  learning-by-confusion scheme for the XY model on a $L=32$ lattice. Also shown are the accuracies from the
threshold-value classification  based on the magnetization $|m|$ and the configurational energy $E$. 
The vertical line denotes the exakt   transition temperature $T_\mathrm{KT}$. 
}
\label{fig_cs_xy_a}
\end{figure}

We may now return to the XY model. 
In  Fig.~\ref{fig_cs_xy_a}, we compare the test accuracy of the learning-by-confusion scheme with the above CNN  to the  threshold-value classification   based on $|m|$ and $E$. There are several points to be noticed here: (i) the test accuracy for the CNN is rather shallow in the low-temperature regime, and it is thus difficult to  identify a clear maximum in the test accuracy. This observation was also made in Ref.~\onlinecite{Beach17}, where the learning-by-confusion scheme was applied to the XY model using a different neural network design, (ii) in the low-temperature regime, the test accuracy of the CNN tends to follow  the threshold-value classification based on the magnetization $|m|$, while it deviates from its more pronounced suppression for larger $T$, (iii)  the accuracy of the  threshold-value classification  based on the energy $E$ is higher than the one  based on $|m|$. It also shows a more shallow overall behavior, and  -- up to a rescaling factor --  traces  the overall shape of the  test accuracy of the CNN .  This observation is in accord with the findings in the previous section: the filters of the CNN provide a (branch-cut corrected) estimate of the local gradients in the spin configuration. These local differences enter  the calculation of the configurational energy  through the cosine functions in $H$. While a spatial average of the local angle differences  provides a gross estimate of the configurational energy,  it is less accurate for the classification process than the actual energy.
We think that for this reason, the test accuracy of the CNN  traces the shape of the threshold-value classification accuracy based on $E$, but falls below its higher accuracy.

Furthermore, the classification of the neural network after training on a given value of $T^*$ is based on the ratio $R=y_1/y_2$ of the activities on the output layer, which depends via  Eq.~(\ref{eq:outratio}) on
the activities of the fully connected layer of the CNN in the exponents. If the later indeed relate to a physical parameter that the neural network has learned,   
then the output  ratio $R$  should reflect the temperature dependence of this  parameter.  In particular, one would expect
 the ratio $R$  to exhibit an enhanced temperature dependence where the physical parameter shows a maximum change with temperature. 

\begin{figure}[t]
\includegraphics[width=\columnwidth]{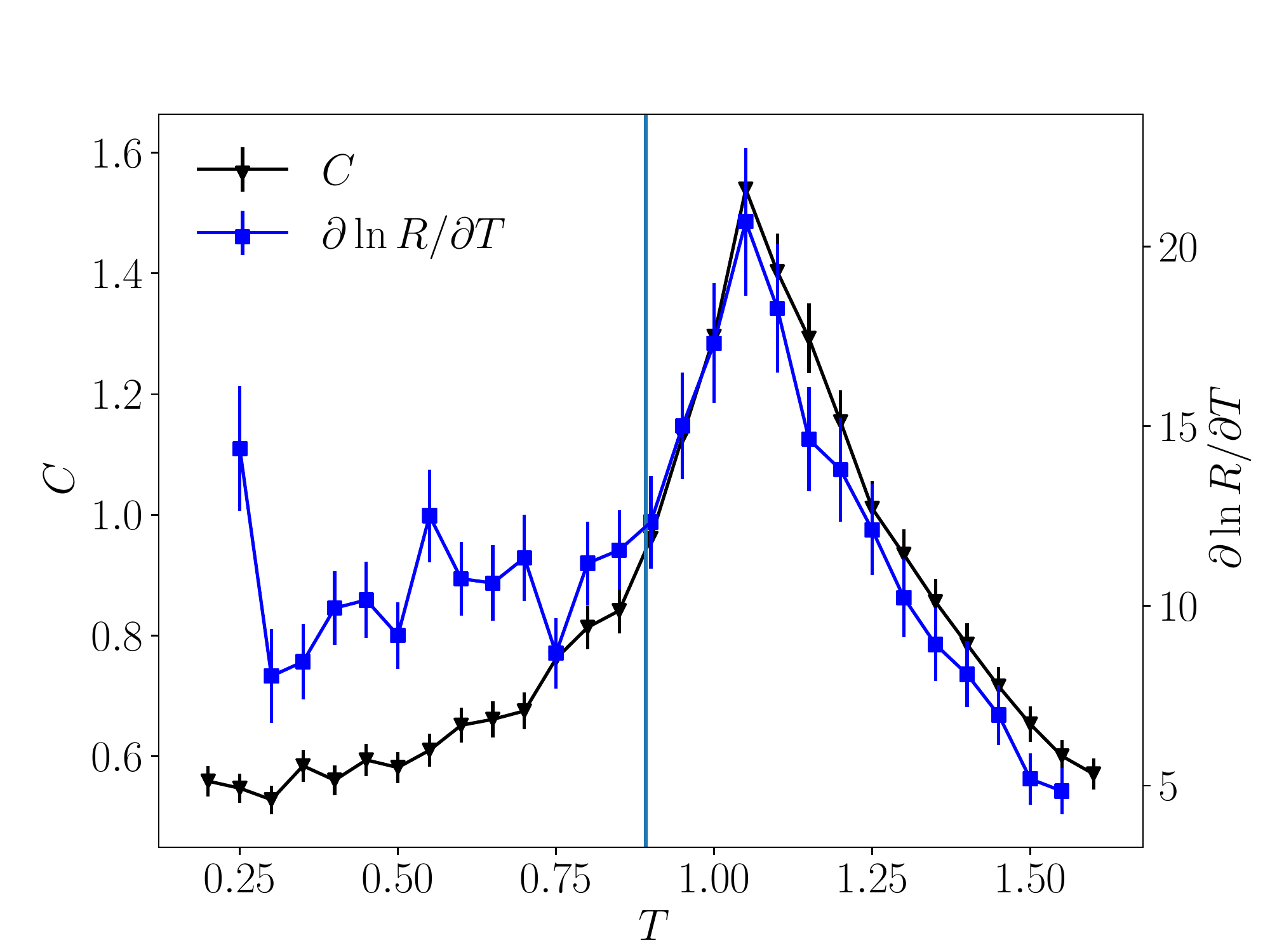}
\caption{(Color online). Logarithmic derivative $\partial \ln R/\partial T$ of the output layer activity ratio $R$ as a function of $T$, compared to the $T$-dependence of the specific heat $C$ for the XY model for $L=32$. 
For the calculation of  $\partial \ln R/\partial T$, the logarithmic derivates were averaged over the considered range of $T^*$-values and input configurations. 
The vertical line denotes the exakt   transition temperature $T_\mathrm{KT}$. 
}
\label{fig_cs_xy_peak}
\end{figure}

Based on this argument, we thus compare in Fig.~\ref{fig_cs_xy_peak} the logarithmic derivative $\partial \ln R/\partial T$  of the ratio $R$  with respect to $T$, averaged over the considered range of $T^*$-values and input configurations,
 to the temperature dependence of the  specific heat of the XY model, $C= \partial (E/N) / \partial T$, which quantities the change in the energy $E$ with temperature.
 We observe a
clear correlation between the behavior of the  logarithmic derivative of the ratio $R$ and the  maximum in the specific heat at $T_\mathrm{max}$.
This adds further support to the previous conclusion, that the configurational energy, which relates to the local angle differences in the XY model, is a relevant 
quantity for the classification process of the trained CNN model.
While at many phase transitions, such as for the Ising model, the specific heat peak, indicating the maximum change in the energy,  indeed 
coincides with the phase transition temperature, this is however not the case for the XY model, where $T_\mathrm{max}$ lies somewhat above $T_\mathrm{KT}$, as mentioned already. 
This shows that a neural network, when trained on the bare spin configurations of the XY model, may not necessarily allow to identify the true transition temperature in the learning-by-confusion scheme.

\begin{figure}[t]
\includegraphics[width=1.1\columnwidth]{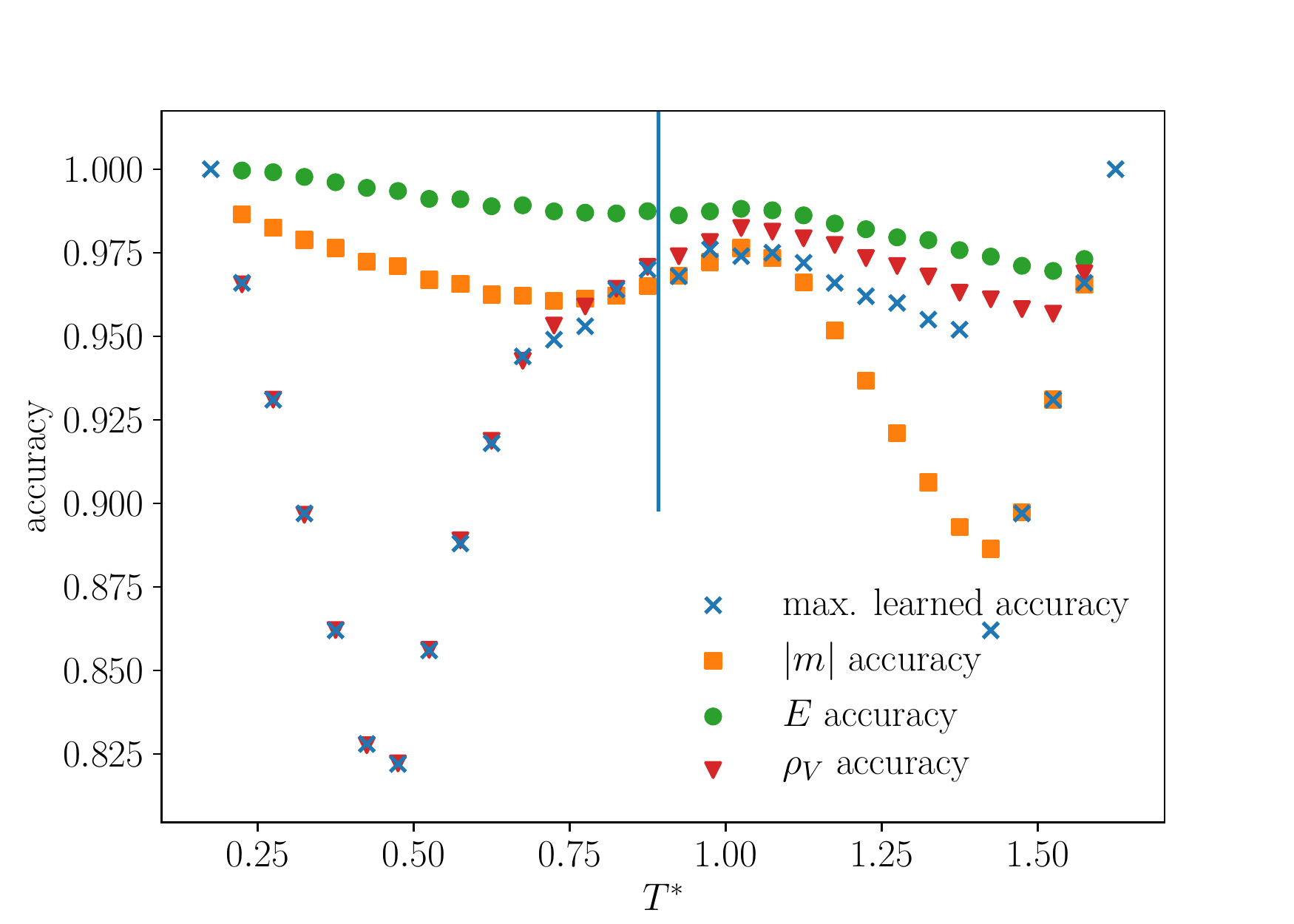}
\caption{(Color online). Test accuracy of the  learning-by-confusion scheme for the XY model on a $L=32$ lattice based on  the vortex configurations. Also shown are the accuracies from the
threshold-value classification based on the magnetization $|m|$, the configurational energy $E$ and the vortex density $\rho_V$. 
The vertical line denotes the exakt   transition temperature $T_\mathrm{KT}$. 
}
\label{fig_cs_xy_v}
\end{figure}

\begin{figure}[t]
\includegraphics[width=\columnwidth]{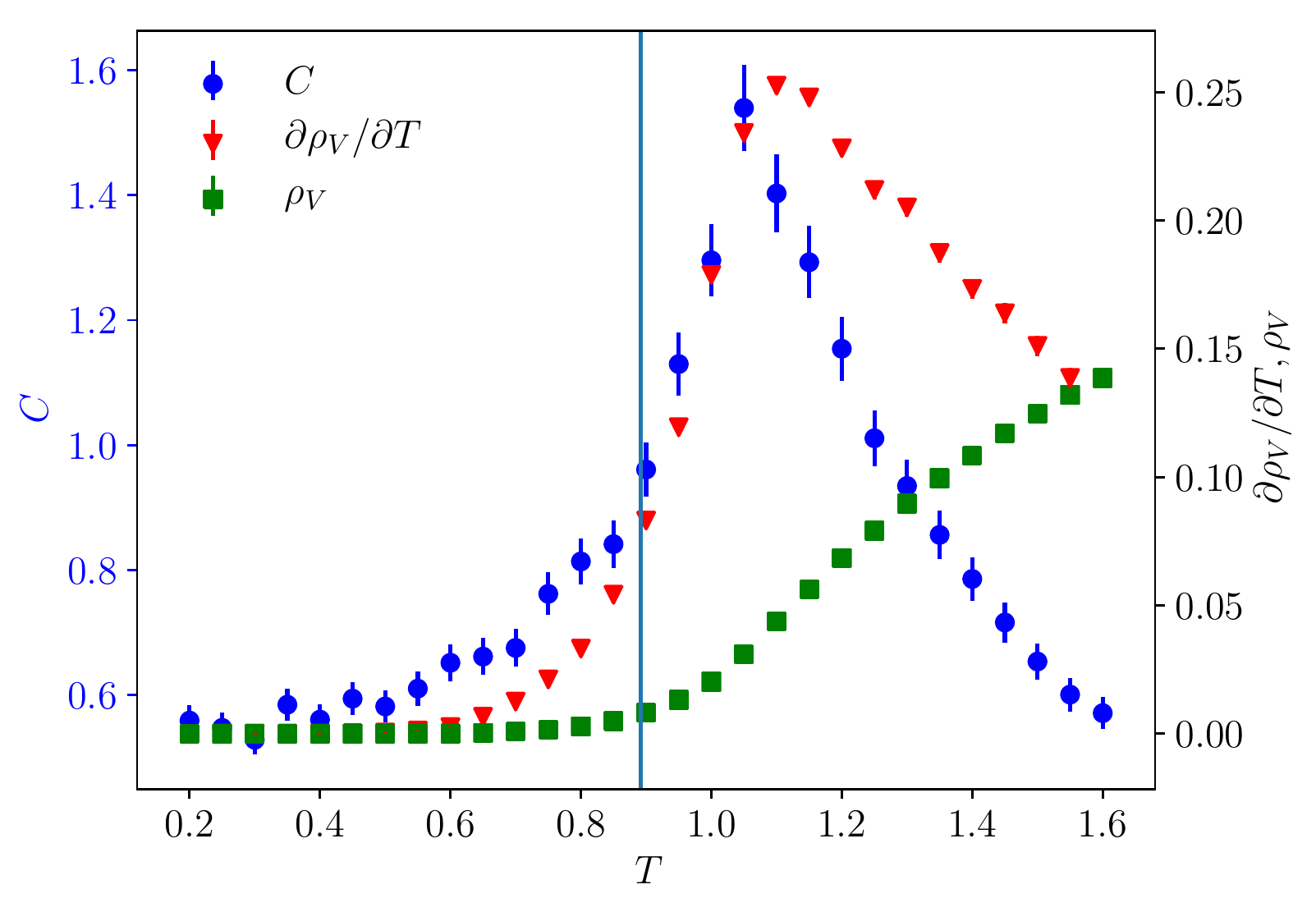}
\caption{(Color online). Temperature dependence of the  vortex density $\rho_V$, its derivative $\partial\rho_V/\partial T$ and the specific heat $C$ for the XY model on the $L=32$ lattice.  
The vertical line denotes the exakt   transition temperature $T_\mathrm{KT}$. 
}
\label{fig_xy_v}
\end{figure}

In that case one may think that a more direct access to the actual physics will be feasible if instead of the spin configurations, one feeds the local vorticities to the input layer. 
Here, we use the following procedure to identify for each plaquette of the square lattice if a local vortex core is present:
Denoting the four spins at the corners of a plaquette $p$ (in anti-clockwise order) as $\phi_{p,1},...,\phi_{p,4}$, we calculate the differences along each edge of the considered plaquette, $\Delta\phi_{p,1}=\phi_{p,2}-\phi_{p,1}$, $\Delta\phi_{p,2}=\phi_{p,3}-\phi_{p,2}$, $\Delta\phi_{p,3}=\phi_{p,4}-\phi_{p,3}$,  $\Delta\phi_{p4}=\phi_{p,1}-\phi_{p,4}$, and shift each of these four values to the interval $[-\pi,\pi)$, via adding integer multiplies of $2\pi$ accordingly. We assign a local value of the vorticity $k_p$ to the considered plaquette upon summing the four shifted angle differences $\Delta\phi_{p,i}\in[-\pi,\pi)$ and normalized by $2\pi$, i.e., 
\begin{equation}
k_p=\frac{1}{2\pi}\sum_i \Delta\phi_{p,i}.
\end{equation}
We then used these  plaquette vorticities as the input data to a CNN input layer instead of the bare 
spin configurations. 
The resulting $T^*$ dependence of the test accuracy of the CNN (with the same layout as before) is shown in  Fig.~\ref{fig_cs_xy_v}.  
We obtain a well developed w-shape in this case. 
Similarly to the one reported in Ref.~\onlinecite{Beach17}, the w-shape of the learning-by-confusion scheme in Fig.~\ref{fig_cs_xy_v} is skewed, even though here we used an essentially symmetric temperature region around $T_\mathrm{KT}$.
Also included in this figure are the threshold-value classification accuracies based on $|m|$, $E$, and the mean vortex density 
\begin{equation}
\rho_v=\frac{1}{N_p}\sum_p |k_p|,
\end{equation}
where $N_p$ denotes the number of plaquettes. From Fig.~\ref{fig_cs_xy_v} we see that the test accuracy of the CNN remarkably closely follows  the  threshold-value classification based on the vortex density,  in particular in the low-temperature regime and with a similarly skewed w-shape. This indicates that this physical quantity is  closely related to the parameter that the network has learned. 
In fact, this quantity is readily accessible to the neural network upon averaging the values of $k_p$ from the input layer.
This result appears satisfying from a physical perspective -- even though of course, we did in this way perform quite some preprocessing, guided by  our knowledge of the underlying physics of the model under investigation. However, the peak of the test accuracy is still located above $T_\mathrm{KT}$. We can understand this behavior by examining the temperature dependence of $\rho_v$. This is shown in Fig.~\ref{fig_xy_v} along with its derivative $\partial \rho_v/\partial T$ and the specific heat $C$. In accord with the already mentioned fact that the specific heat peak at $T_\mathrm{max}$ results from an enhanced proliferation of free vortices, we observe that the maximum in $\partial \rho_v/\partial T$ is close to $T_\mathrm{max}$ as well.

We thus find that in both cases, after training the considered CNN on either the  bare spin configurations or on vortices, the learning-by-confusion scheme 
predicts a transition temperature $T^*$ that is set by the value of $T_\mathrm{max}$ instead of the actual transition temperature $T_\mathrm{KT}$, 
due to the enhanced change at  $T_\mathrm{max}$ in  the relevant parameters that the system learns.  Since $T_\mathrm{max}$ remains above  $T_\mathrm{KT}$  in the thermodynamic limit, 
we expect that  this behavior  persists also if much larger system sizes would be considered. 

%

\section{Discussion}
\label{Sec:Discussion}

In the first part of this study, we examined  the classification process of  shallow fully connected and convolutional  neural networks for the Ising model, focusing on the effect of extended domain wall configurations. By including such configurations in the learning batch configurations, the fully connected neural network learned  to identify horizontally and vertically striped domains.  Increasing the number of hidden neurons, the network can locate such  patterns over a larger range of positions and both polarizations.  We found  the convolutional neural network to exhibit two  major classes of filter kernels that either propagate locally averaged values of the magnetization to the fully connected layer or identify local domain walls in the input configuration, which upon summation over the filter positions represent an estimate of the configurational energy. This information is used, along with the magnetization, to obtain a  highly accurate classification process.

In a similar convolutional neural network for the XY model, we  identified filters that  detect local directional differences in the spin configuration, while other  filters apparently correct for   false identifications of large gradients across the branch-cut in the cyclic angle variables. Hence, for this convolutional neural network the configurational energy (or an estimate thereof)  is again a relevant physical quantity for the classification process. Additional insight was  obtained from the learning-by-confusion scheme. Its test accuracy can  be directly compared to a  threshold-value classification method, which we introduced  as a  means of directly assessing the relevance of specific physical observables for the network's classification process.
For the XY model, we obtained in this way additional evidence for the relevance of the local angle gradients for the classification process of the considered convolutional neural network. 
Upon examination of the  derivative of the output level activity ratio, we noticed a strong correlation with the specific heat peak. 
This allowed us to extract a corresponding  temperature value,  even though the test accuracy of the learning-by-confusion scheme does not exhibit a pronounced w-shape.
However, the specific heat peak of the XY model does not signal the actual transition temperature, but is located  above $T_\mathrm{KT}$. This particular property of the XY model keeps
the learning-by-confusion scheme based on the considered convolutional neural network  from identifying the actual transition temperature.

A neural network may thus be able to perform the classification task with a high   accuracy based on a (physical) quantity, but  this quantity need  not  relate in the anticipated way to the actual phase transition. Of course, such issues may 
depend in a delicate way on the network design and could possibly be avoided by appropriately preprocessing the bare model configurations before feeding them to the network. 
In this respect, we however noticed that for the XY model the situation was not improved upon by feeding  the vortex configurations to the input layer. 
In this case, the network  readily learned the vortex density, but the temperature-dependence of this quantity also does not  identify $T_\mathrm{KT}$, since  the most pronounced \textit{change} in the vortex density is due to an enhanced vortex proliferation, corresponding to the specific heat peak. 
As a generic tool to locate phase transitions such schemes may thus be  difficult to control, which would be  an issue
 in view of  models for which the underlying physics is not  that  well understood yet.

On the other hand, here we focused our diagnostic approach on rather shallow neural networks, for which we could readily examine the inner structure  in terms of weight matrices and a small number of filter kernels. 
Even though the classification performance of these shallow networks proved to be high,  it may still  be expected -- given the relation to the renormalization group~\cite{Beny13,Mehta14} -- that deep learning networks, based on several convolutional layers and a more complex network layout allow for  (i) a hierarchy of physical parameters for the classification process  to emerge on increasing length scales, and thus (ii) a  higher level of robustness with respect to the above mentioned issues.  
However, for the XY model, Ref.~\onlinecite{Beach17} observed that a multi-layer convolutional neural network with an optimal design to identify vortices from the  bare spin configuration is only a locally stable  solution of the learning procedure. 
Monitoring the  derivative of the output activity ratio  and the threshold-value classification scheme can of course also be applied to such multi-layer network, as well as to
other complex neural  networks and may thus be useful for further assessments of machine learning methods for condensed matter theory research.

\section*{Acknowledgments}
We thank  P. Emonts, S. Hesselmann, T.C. Lang, Z. Y. Meng, and L. Wang  
for useful discussions
and acknowledge support by the Deutsche Forschungsgemeinschaft (DFG) under grant FOR 1807 and RTG 1995. Furthermore, we thank the IT Center at RWTH Aachen University and the JSC J\"ulich for access to computing time through JARA-HPC.

\end{document}